\documentclass[final,5p,times]{elsarticle}
\usepackage{mathtools}
\usepackage{amsmath,amsthm,amssymb}
\usepackage{wasysym}
\usepackage{graphicx}
\usepackage{wrapfig}
\usepackage{caption}
\usepackage{lineno}
\usepackage[unicode]{hyperref}
\usepackage{mdframed}
\usepackage{miller}
\usepackage{natbib}

\newcommand{\rev}[1]{#1}

\begin{document}

\journal{Nuclear Instruments and Methods in Physics Research Section A}

\begin{frontmatter}

\title{Development of Ultrafast and Radiation-Hard GAGG for the Next-Generation of High-Energy Physics Calorimeters\tnoteref{t1}}
\tnotetext[t1]{This manuscript is currently under revision at \emph{Nuclear Instruments and Methods in Physics Research Section~A}.}

\author[cern,ilm]{Louis Roux}
\author[cern]{Loris Martinazzoli}
\author[cern,stras]{Julie Delenne}
\author[cern]{Philipp Roloff}
\author[crytur,fzu]{Ondřej Zapadlík}
\author[crytur]{Jan Polak}
\author[crytur]{Jan Havlíček}
\author[crytur]{Silvia Sýkorová}
\author[fzu]{Martin Nikl}
\author[fzu]{Pavel Boháček}
\author[ilm,iuf]{Christophe Dujardin}
\author[cern]{Etiennette Auffray}

\affiliation[cern]{organization={European Organization for Nuclear Research (CERN)},%
            city={Geneva},
            country={Switzerland}}
\affiliation[ilm]{organization={University Claude Bernard Lyon1, CNRS, ILM UMR5306},
            city={Lyon},
            country={France}} 
\affiliation[stras]{organization={University of Strasbourg, CNRS, IPHC UMR7178},
            city={Strasbourg},
            country={France}} 
\affiliation[crytur]{organization={Crytur, Ltd},
            city={Turnov},
            country={Czech Republic}}  
\affiliation[fzu]{organization={Institute of Physics of the Academy of Sciences of the Czech Republic (FZU)},
            city={Prague},
            country={Czech Republic}}  
\affiliation[iuf]{organization={Institut Universitaire de France (IUF)},
            city={Paris},
            country={France}}

\begin{abstract}
The evolution of High Energy Physics (HEP) toward future collider experiments with High Luminosity (HL), such as the HL-LHC, requires the development of scintillating materials that combine high density, excellent radiation hardness and an ultrafast response. While Cerium-doped Gadolinium Aluminum Gallium Garnet (GAGG:Ce) offers a very high light yield and resilience to irradiation, its typical decay time of approximately 50--60 ns may lead to pile-up effects in high-rate environments.

In this paper, we report on the development and multi-stage characterization of various accelerated GAGG compositions optimized for timing performance and grown by Crytur. By taking advantage of divalent co-doping to engineer the scintillation kinetics, we achieved an effective decay time ($\tau_{\mathrm{d,eff}}$) down to 5.5 ns while maintaining a competitive light yield of several thousand photons per MeV.

Laboratory characterization demonstrates that the time resolution under gamma-ray excitation is comparable to commercial GAGG, while the time resolution measured with 120 GeV pions reaches performance levels comparable to state-of-the-art LYSO:Ce,Ca. \rev{After a 1~MGy proton irradiation campaign the material retains most of its optical transmission.}
The results confirm that this ultrafast GAGG composition is a viable candidate for the next generation of HEP calorimetry and timing detectors.

\end{abstract}

\begin{keyword}

High-Energy Physics \sep Calorimetry \sep Advanced scintillating materials \sep Fast timing \sep Radiation detection 
\end{keyword}

\end{frontmatter}

\section{Introduction}
The upcoming High Luminosity phase of the Large Hadron Collider (HL-LHC) at CERN, along with future high-energy collider projects \cite{FCC}, presents an unprecedented challenge for radiation detection technology. Detectors in these environments must withstand extreme ionizing doses; for instance, up to 1 MGy is expected in the inner region of the electromagnetic calorimeter of LHCb \cite{LHCb_ftdr, LHCb_scoping}, while maintaining high-performance time resolution. To meet these stringent requirements, developing inorganic scintillators that combine high density, radiation hardness, and ultrafast timing response is crucial for building high-performance calorimeters with excellent energy and time resolution.

Among the candidates for future calorimetry and timing layers, multi-component garnet crystals, specifically Cerium-doped Gd$_3$Ga$_x$Al$_{5-x}$O$_{12}$ ($x \in [2.5,\,3]$) Gadolinium Aluminum Gallium Garnet (GAGG:Ce) have emerged as a leading solution. Since their first reported combinatorial studies in \cite{kamada_2011}, GAGG:Ce has gained significant interest due to a light yield that can exceed three times that of classical scintillators like YAG:Ce or LuAG:Ce. Its high density ($6.6$ g/cm$^3$) and short radiation and nuclear interaction lengths ($X_0$ = $1.6$ cm and $\lambda_\mathrm{int}$ = $22.4$ cm, respectively) make GAGG suitable for compact calorimetry \cite{Martinazzoli2021}.

Furthermore, its non-hygroscopic nature and ease of machining allow for the creation of complex detector geometries, such as the high-aspect-ratio crystal fibers used in \textit{spaghetti} calorimeters \cite{AN2023167629}.

Despite these advantages, non-ultrafast (non-accelerated) GAGG faces a critical bottleneck in high-rate environments. This term is used here for both standard GAGG:Ce and the faster commercial grades derived from it, such as GFAG, whose decay times remain of several tens of nanoseconds. The LHC proton beams are split into bunches that collide every 25 ns. Non-ultrafast GAGG typically exhibits a relatively slow primary scintillation decay component (50--60 ns). This leads to ``pile-up,'' where signals from successive bunch crossings may overlap, degrading both energy reconstruction and time resolution. To operate effectively at the HL-LHC, the scintillation response must be accelerated to ensure clear signal separation and reject background noise.

Recent breakthroughs in material engineering have focused on band-gap and defect engineering \cite{KAMADA201563, LUCCHINI2016176, Martinazzoli2022} to overcome these timing limitations. By co-doping GAGG:Ce with divalent ions such as Mg$^{2+}$, researchers have successfully stabilized Cerium in the Ce$^{4+}$ charge state. These Ce$^{4+}$ ions act as efficient competitors with electron traps for the immediate capture of electrons from the conduction band, increasing the portion of prompt radiative electron-hole recombination which accelerates scintillation response. Heavy doping and codoping by Ce and Mg in GAGG even accelerates the photoluminescence and scintillation decay well below the photoluminescence lifetime of Ce$^{3+}$ (60 ns) down to several nanoseconds due to quenching process in closely spaced Ce-Mg pairs. This acceleration comes at the expense of a decreased light yield. The resulting reduction in decay time down to a few nanoseconds is nevertheless a necessary trade-off for High-Energy Physics (HEP) applications where timing is critical.
Building on earlier proof-of-concept work with accelerated compositions \cite{Martinazzoli2022}, this study advances toward mass production by developing inorganic scintillators that combine high density, radiation hardness, and ultrafast timing response crucial for achieving precise time resolution.

We present a comprehensive study beginning with the material synthesis and laboratory characterization, including the investigation of optical, timing, and radiation-tolerance
measurements. Finally, we validate the performance of these crystals under realistic experimental conditions through test beam measurements, evaluating their time resolution when subjected to 120 GeV pions and electrons. These results confirm the potential of accelerated GAGG to meet the rigorous demands of next-generation particle physics experiments.

\section{Experimental Samples and Material Properties}
\subsection{Crystal samples}

Four distinct ingots were grown for this study: three primary ingots (A, B, and C) and one additional experimental ingot (D). It is included only to extend the light-yield/decay-time trend of Section~\ref{sec:emiss}; the available samples are those listed in Table~\ref{tab:gagg_ingots_comp}. From the grown ingots, samples were extracted from both the head and tail sections, allowing for a direct comparison of their optical and scintillation properties along the growth axis. To ensure optical clarity and optimal light collection, all samples were fully polished on all surfaces prior to characterization. Specific geometries were cut from ingots A, B, and C to facilitate different measurement types $2 \times 2 \times 3$~mm$^3$ (pixels), $2 \times 2 \times 10$~mm$^3$, $1 \times \diameter 10$~mm$^3$ (plates), $10 \times 10 \times 10$~mm$^3$ and crystal fibers: $1 \times 1 \times 50$~mm$^3$ and $1 \times 1 \times 100$~mm$^3$. In this article, ``crystal fiber'' denotes these high-aspect-ratio polished crystal bars (no cladding or optical-waveguiding structure), in the sense used for \emph{spaghetti} calorimeters.

\subsection{Crystal growth procedure}
GAGG:Ce,Mg single crystals were grown by the Czochralski (CZ) technique using radiofrequency (RF) inductive heating. The growth process was carried out in an iridium crucible under a controlled inert atmosphere to prevent oxidation and ensure chemical stability at elevated temperatures. High-purity starting materials (4N, 99.99\%) were used to prepare the melt, which was homogenized prior to seeding to ensure uniform dopant distribution.
A GAGG crystal oriented along the \hkl<1 0 0> direction served as the seed. Crystal growth was performed under carefully controlled thermal conditions to maintain a stable solid–liquid interface and minimize defect formation. The growth parameters, including pulling rate and rotation speed, were optimized to promote high optical quality and compositional homogeneity. Particular attention was paid to maintaining stable thermal gradients to suppress the formation of inclusions and dislocations.
The resulting cylindrical ingots had a diameter of approximately 40 mm. Standard crystals reached a length of about 55 mm, whereas ingot B was grown to an extended length of 105 mm. Ingots A and B exhibit similar cerium concentrations, while ingot C contains approximately 20\% lower cerium content.
The magnesium concentration was analyzed using electron probe microanalysis (EPMA) and glow discharge mass spectrometry (GDMS), revealing a homogeneous distribution throughout the ingots. An exception was observed for ingot B, where a variation of approximately 15\% between the top and bottom sections was detected. This ingot represents the first growth attempt with an extended crystal length of 105 mm, carried out under non-optimized conditions, which likely accounts for its reduced compositional homogeneity. In addition, crystal B exhibited a significant number of inclusions, likely originating from trapped gas bubbles or metallic impurities.

\begin{table*}[ht]
    \centering
    \caption{Inventory of the GAGG samples available from each ingot, with dimensions in mm$^3$.}
    \label{tab:gagg_ingots_comp}
    \begin{tabular*}{\textwidth}{@{\extracolsep{\fill}}lcccccc}
        \hline
        \hline
        Ingot & $2\times2\times3$ & $2\times2\times10$ & $10\times10\times10$ & $1\times1\times50$ & $1\times1\times100$ & $1 \times 10$ (\diameter) \\
        \hline 
        A & X & X & & & & X \\
        B & X & & & & & X \\
        C & X & & X & X & X & X \\
        D & & & & & & X  \\
        \hline
        \hline
    \end{tabular*}
\end{table*}

\section{Methods and Instrumentation}
All characterization measurements were performed at CERN. The test benches, sample preparation procedures, and measurement methods are described in the following subsections.

\subsection{Transmission and absorbance}
The transmission spectra were measured using a Perkin Elmer Lambda 650 UV/VIS spectrophotometer. The instrument is provided with two lamps, a halogen-tungsten one operating between 314 and 900 nm and a deuterium lamp working down to 190 nm. Their light passes through a monochromator and then splits into two branches. One of the monochromatic light beams traverses the crystal sample placed on a moving stage. The second beam serves as a reference channel to monitor and correct for instrumental drift throughout the data acquisition process. Both beams are focused onto a photomultiplier tube for intensity measurement \cite{robertothesis}. The transmittance $T$ is then determined from the ratio of the two beam intensities, as defined in Eq.~\ref{eq:transmission}, and the absorbance $A$ as defined in Eq.~\ref{eq:absorbance}.

\begin{equation}\label{eq:transmission}
    T(\lambda) = \frac{I(\lambda)}{I_0(\lambda)}
\end{equation}

\begin{equation}\label{eq:absorbance}
    A(\lambda) = -\log_{10}T(\lambda)
\end{equation}

Where $I(\lambda)$ is the intensity transmitted through the sample and $I_0(\lambda)$ the incident (reference-beam) intensity, both measured as a function of the wavelength $\lambda$.

\subsection{Radiation hardness}\label{sec:rad_meth}
As part of a radiation hardness investigation, the $10\times10\times10$ mm$^3$ sample underwent proton irradiation at the PS-IRRAD Proton Facility (CERN) with 24~GeV protons at a fluence $\approx 4.0\cdot10^{15}$ protons$\cdot$cm$^{-2}$ which is equivalent to a radiation dose of 1 MGy. Transmission measurements were then performed roughly 6 months after exposure, allowing for a cool down period, to detect any discernible and measurable changes.

\begin{equation}\label{eq:ind_abs}
    \mu_{\mathrm{ind}}(\lambda) \coloneqq \frac{1}{L} \cdot \ln \frac{T_{\mathrm{before}}(\lambda)}{T_{\mathrm{after}}(\lambda)}
\end{equation}

To quantify the loss of transmission from irradiation we computed the radiation-induced absorption coefficient ($\mu_{\mathrm{ind}}$) which is obtained as defined in Eq.~\ref{eq:ind_abs} where $L$ is the thickness of the sample being traversed (here $L$ = 10 mm) and $T_{\mathrm{before/after}}$ are the transmission before and after irradiation of the measured sample under study.

\subsection{Photoluminescence and radioluminescence spectra}

The emission spectra of photoluminescence (PL) were measured with a PerkinElmer LS55 spectrofluorometer. The spectrofluorometer emits a tunable monochromatic beam of light onto the sample and measures the emitted light intensity at 90° as a function of wavelength, a configuration referred to as reflection mode. In order to identify the optimal spectral parameters, preliminary excitation and emission scans were performed on the samples. The emission spectra of radioluminescence (RL) were measured with an Avantes spectrometer (AvaSpec-HSC1024x58TEC-EVO), using an Amptek Mini-X X-ray tube as the excitation source (operating voltage: 40 kV), in transmission mode. Both measured spectra were corrected for experimental distortions using the built-in calibration curves provided by the respective manufacturers.

\subsection{Light attenuation}\label{sec:att_meth}
\begin{figure}[!ht]
    \centering
    \includegraphics[width = .9\linewidth]{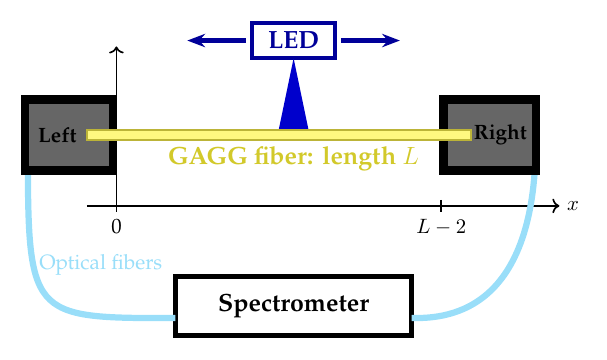}
    \caption{Schematic of the attenuation bench working principle.}
    \label{fig:att_bench}
\end{figure}

The setup is as follows: an LED is used to excite the GAGG crystal fiber at 445 nm in various position along the longitudinal direction of the latter. 
Light collection at both ends of the crystal fiber is facilitated by cosine correctors. The end faces of the crystal fiber were in contact with cosine correctors mounted on optical fibers bringing light to Avantes.
The collected light is then directed into an Avantes AVASPEC-ULS2048CL-2-RS-EVO spectrometer. The setup is illustrated in Fig.~\ref{fig:att_bench}. Due to mechanical limitations, 1 cm at each end of the crystal fiber is excluded.

\subsection{Light yield}\label{sec:ly_meth}

Each sample was wrapped with several layers of Teflon to maximize light collection and coupled to a Hamamatsu R2059 PMT using Rhodorsil 55 optical grease ($n$ = 1.41), then excited using a $^{137}$Cs source emitting 661.7 keV $\gamma$-rays. The PMT was connected to a DT5720 CAEN digitizer operating in charge integration mode, with an analog attenuator available to prevent signal saturation. To protect the PMT from external light, both the PMT and samples were covered with a light-tight black paper lid, and the entire setup was enclosed in a temperature-controlled black box. Conversion to photons was obtained by measuring the single photo-electron peak of the PMT and correcting for its quantum efficiency, weighted by the emission spectra of the GAGG samples.

\subsection{Scintillation kinetics}
The scintillation emission kinetics of the GAGG samples were measured using a Time Correlated Single Photon Counting (TCSPC) setup, described in \cite{GUNDACKER201842} and \cite{Pagano}. A pulse diode laser (PicoQuant (LDH-P-C-405) with driver PDL 800-B) was used as an excitation source of an X-ray tube XRT N5084t from Hamamatsu. It generated an X-ray beam with a continuous energy spectrum between 0 and 40 keV (with a $\simeq$ 10 keV mean energy). The beam was collimated on the tested sample and its scintillation light was collected using a hybrid photomultiplier (HPM 100-07 Becker Hickl).
A 400 nm long pass filter was placed before the hybrid photomultiplier to reduce background. 
The measurements were done by exciting the sample on one surface and detecting the emitted light from the same surface (reflection mode). The signal of the hybrid PMT was then fed to an amplifier and timing discriminator and was then used as the stop signal of a time-to-digital converter, while the start was provided by an external trigger of the pulse diode laser.
\begin{equation}\label{eq:scinti_eq_bso}
f(t\mid\theta)=\sum_{i=1}^{3}\,\rho_{i}\cdot\frac{\exp\Bigl(-\frac{t-\theta}{\tau_{\mathrm{d},i}}\Bigr)-\exp\bigl(-\frac{t-\theta}{\tau_{\mathrm{r},i}}\bigr)}{\tau_{\mathrm{d},i}-\tau_{\mathrm{r},i}}\cdot\Theta(t-\theta)+ \varepsilon
\end{equation}

Each scintillation pulse obtained for the samples was modeled as described in Eq.~\ref{eq:scinti_eq_bso}.
The $\tau_{\mathrm{r},i}$ and $\tau_{\mathrm{d},i}$ are the $i$-th components of the rise and decay time constants respectively, while $\rho_i$ is the weight of the i-th component, $\theta$ is the instant at which the scintillation pulse starts and $\varepsilon$ is the background. The scintillation pulse function defined above was then convoluted with the Impulse Response Function (IRF) of the setup ($\simeq$ 121 ps FWHM) to fit the decay time spectra obtained.

From the decay time constants obtained and their weights, the effective decay time of the sample can be evaluated as given in Eq.~\ref{eq:taueff}, where $n$ is the number of components (3 exponentials will be used for the GAGG characterized in this article) used for the fit. 

\begin{equation}\label{eq:taueff}
    \tau_{\mathrm{d,eff}}^{-1} = \sum_{i=1}^{n=3} \frac{\rho_i}{\tau_{\mathrm{d},i}}
\end{equation}

We also compute the weighted average decay time, where each decay component is weighted by its relative intensity, and the time constant $\tau_{1/e}$ which is defined as the time it takes for the scintillation pulse to decay by a factor of $ \frac{1}{e}$.

\subsection{Time resolution}
Precision timing has become a cornerstone of modern detector design, essential for mitigating pile-up in High Luminosity (HL) environments but also for medical applications such as Time-of-Flight Positron Emission Tomography (ToF-PET) \cite{Conti2019}. To fully evaluate the timing performance of the accelerated GAGG compositions, we conducted measurements using two different setups across two distinct excitation regimes. These are described in the following subsections.

\subsubsection{Coincidence time resolution - CTR}\label{sec:ctr_setup}
The CTR of the samples tested was measured with the test set-up using High-Frequency (HF) readout as described in \cite{Gundacker2019}.

Each sample was wrapped in several layers of Teflon coupled to
a NUV-HD-MT (Broadcom) SiPM, using the Cargille Meltmount optical glue ($n$ = 1.539).

To evaluate the CTR, only events from the photopeak were considered. The photopeak window was set  
with a lower threshold at 470 keV and an upper threshold at 555 keV. Considering the events extracted from photopeak selection, the distribution of the time delay between the left and right sample timestamps is produced. 

Thus, the CTR FWHM ($\simeq 2.35 \cdot \sigma$) of the GAGG samples was evaluated from a Gaussian fit of the energy-filtered time delay distribution. 

A scan of the Leading Edge Threshold (LET) on the time signal (to find the optimal timestamp) was performed sample by sample powering their SiPMs at 48~V when using the $3.72\times3.62$~mm$^2$ NUV-HD-MT.
To find the optimal value one can then plot the CTR FWHM for every LET values. The data points were then fitted using Eq.~\ref{eq:ctr_min}.

\begin{equation}\label{eq:ctr_min}
     f(x \mid a_0, a_1, a_2) = \sqrt{a_0^2 + (a_1 \cdot x)^2 + \left( \frac{a_2}{x} \right)^2}
\end{equation}

For the CTR analysis presented hereafter, the reported values correspond to the minimum resolution extracted from the fitting procedure.

\subsubsection{Time resolution with 120 GeV pions}
\begin{figure}[!ht]
    \centering
    \includegraphics[width = 1.\linewidth]{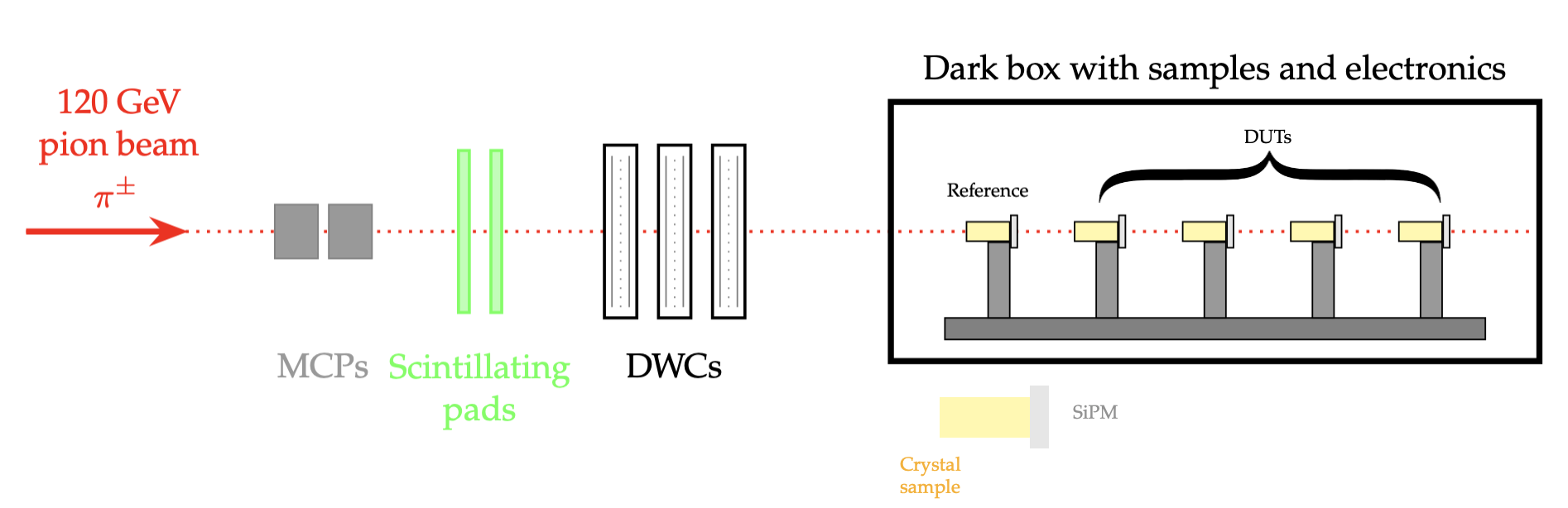}
    \caption{Schematic of the experimental setup for time resolution measurements with 120 GeV pions. The inset provides a detailed close-up of the DUTs and the reference detector, which consists of a crystal coupled to a SiPM.}
    \label{fig:mips_setup}
\end{figure}

The measurements were performed at the CERN SPS H4 beamline in 2024. The beam consisted of 120~GeV pions, whose energy deposition is close to the minimum-ionizing regime.
The testbeam set-up is schematically explained in Fig.~\ref{fig:mips_setup} and the analysis technique is described with more precision in \cite{Cala_2025}.

The hardware trigger was provided by a coincidence of two plastic scintillating pads, while the timing reference was established using two Micro-Channel Plate (MCP) detectors (PMMA Cherenkov radiators attached to MCP-PMTs). For spatial selection and tracking, three Delay Wire Chambers (DWCs) were employed, with the specific selection criteria detailed in the following section. These DWCs operated with an Ar/CO$_2$ gas mixture.

The samples were housed in a light-tight box installed on a motorized XY stage moving in the plane orthogonal to the beam direction. Each crystal-SiPM detector was attached to a HF readout board, similar to the ones described in the section \ref{sec:ctr_setup}, within a metal shielding case. 

\section{Experimental results}
\subsection{Transmission and absorbance}\label{sec:trans}

\begin{figure}[!ht]
    \centering
    \includegraphics[width = 1.\linewidth]{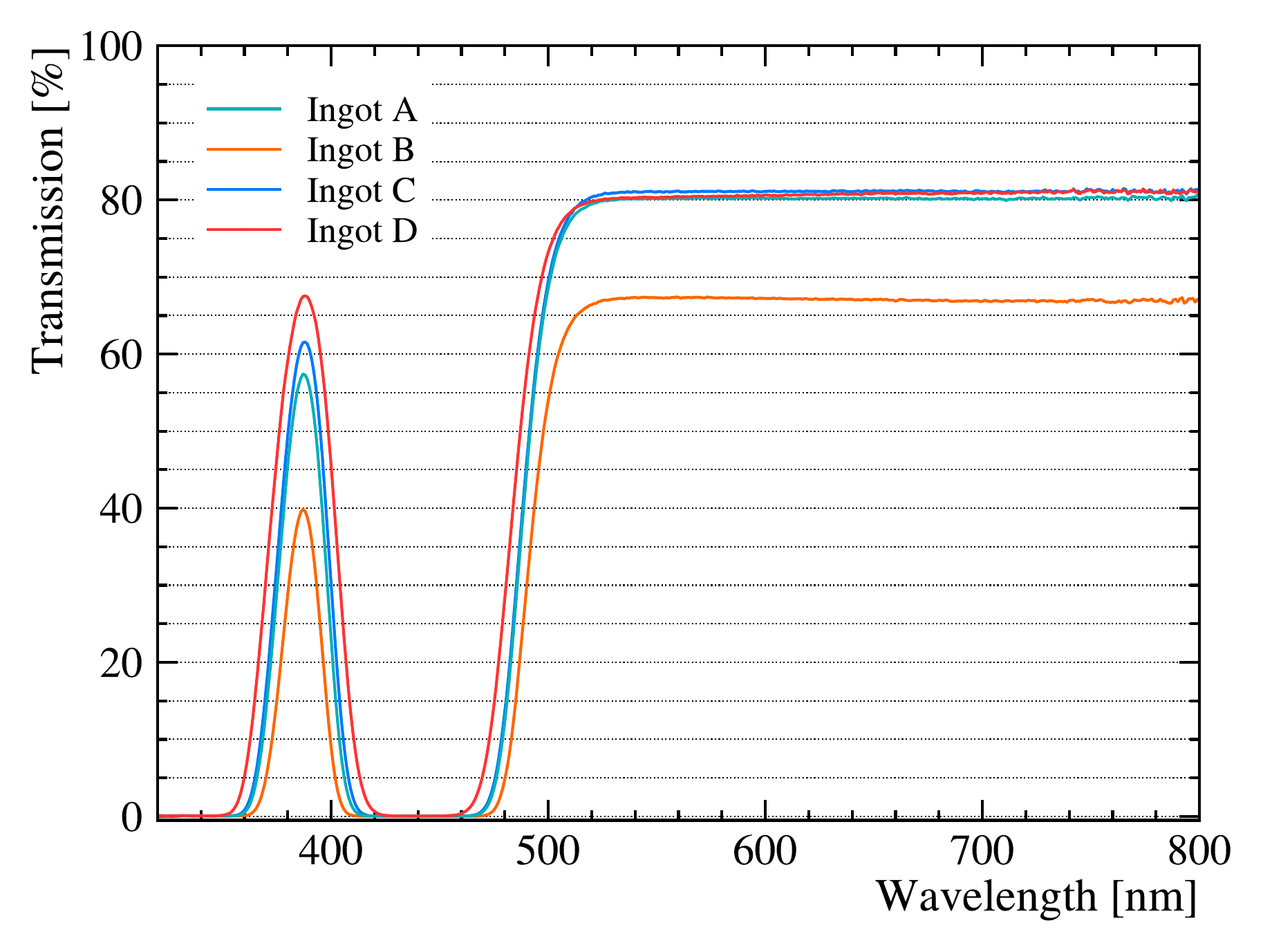}
    \caption{Transmission spectra of the GAGG plates.}
    \label{fig:99_transmi}
\end{figure}

The transmission measurements of the ingots (plate samples) are shown in Fig.~\ref{fig:99_transmi}. 

At the 535 nm emission peak, ingots A, C, and D maintain a transmittance of 80\%, whereas ingot B shows a marked decrease to 67\%. This trend remains consistent in the ultraviolet regime; at 380 nm, ingot B again displays lower transparency than its counterparts. These combined features suggest that fluctuations in dopant distribution and light scattering might compromise the overall optical quality of ingot B.

\begin{figure}[!ht]
    \centering
    \includegraphics[width = 1.\linewidth]{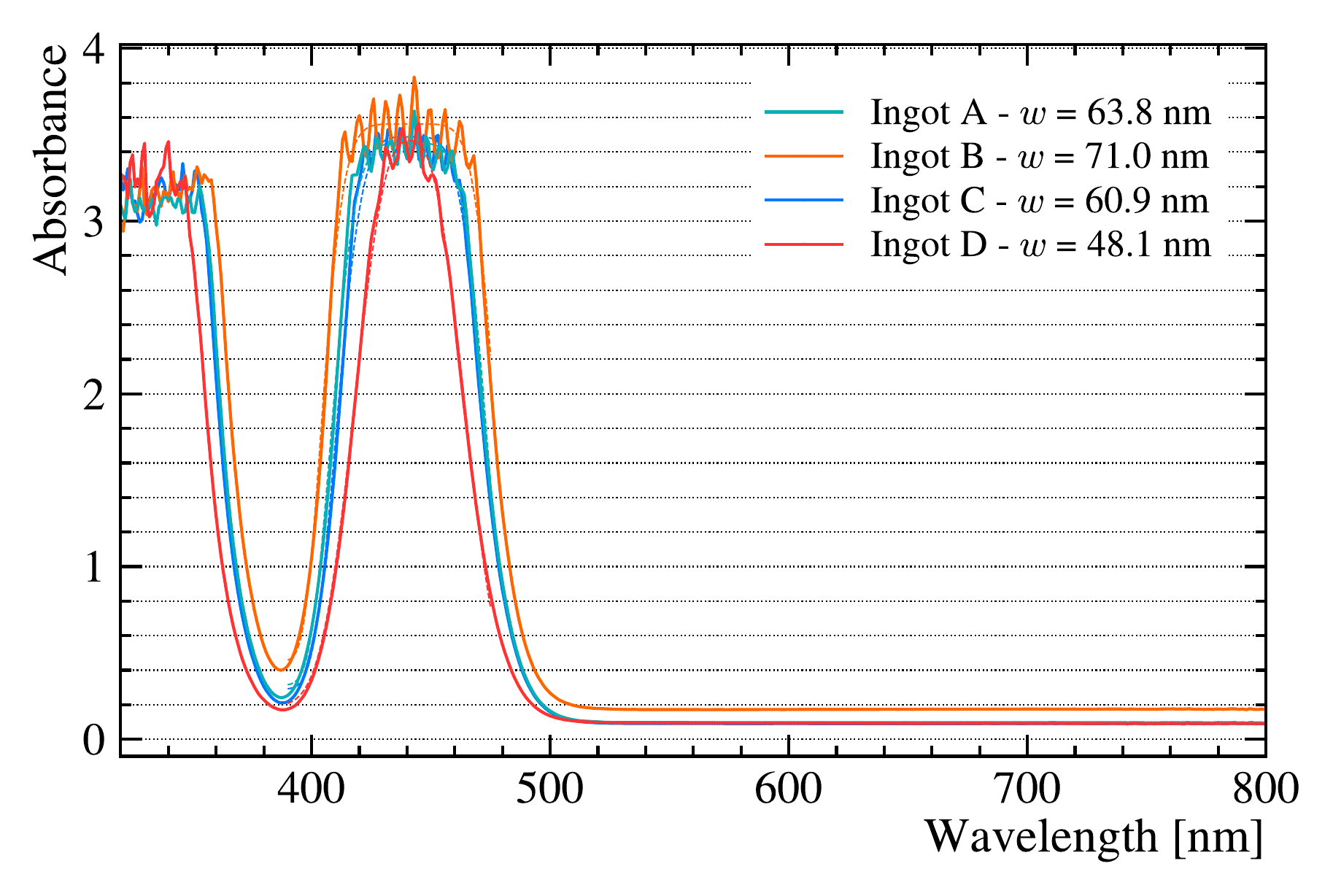}
    \caption{Absorbance spectra of the GAGG plates. The parameter $w$ appearing in the figure legend is the spectral width of the broad absorption band ($\sim$400--500 nm); discussed in Section~\ref{sec:disc}.}
    \label{fig:gagg_absorbance}
\end{figure}

The results are shown in Fig.~\ref{fig:gagg_absorbance}. The acquired spectra reveal the characteristic Ce$^{3+}$ $4f\mapsto5d_{1}$ absorption band at approximately 445 nm, as in \cite{Martinazzoli2021}. The $4f\mapsto5d_{2}$ transition, expected around 330--340 nm, is not resolved, being masked by the absorption cutoff discussed in Section~\ref{sec:disc}.

\subsection{Radiation hardness}\label{sec:rad_res}
Regarding the irradiation results, no visible change in the color of the irradiated 1~cm$^3$ sample was observed, indicating the probable absence of macroscopic radiation-induced color centers.

\begin{figure}[!ht]
    \centering
    \includegraphics[width = 1.\linewidth]{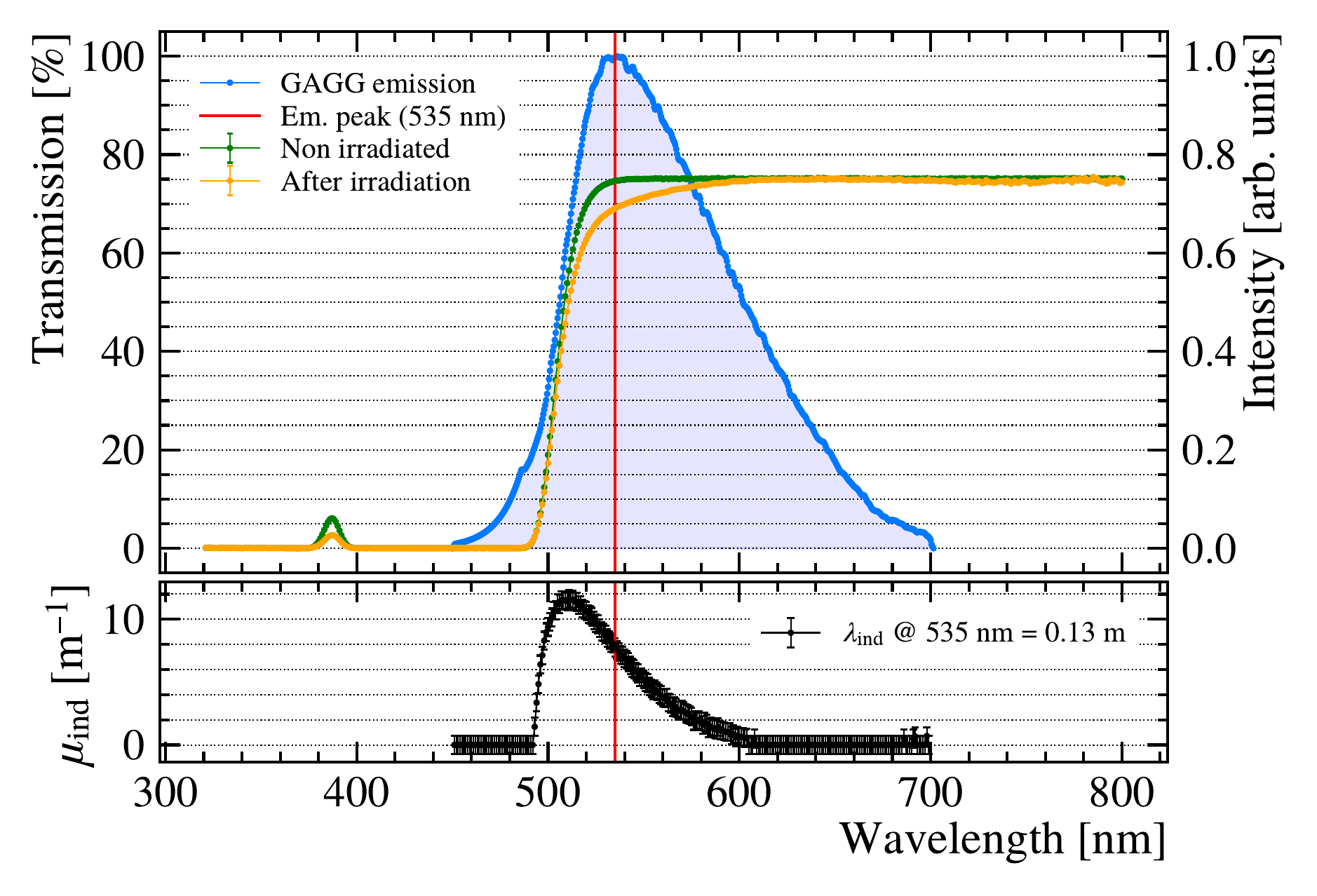}
    \caption{\textit{Top:} Transmission spectra of 1 cm$^3$ GAGG from ingot C before (green) and after (yellow) irradiation with emission spectrum. \textit{Bottom:} Induced absorption coefficient ($\mu_{\mathrm{ind}}$) as a function of the wavelength computed from transmission measurements.}
    \label{fig:gagg_irrad}
\end{figure}

Fig.~\ref{fig:gagg_irrad} presents the radiation-induced optical changes: the upper panel shows transmission spectra before and after irradiation with the GAGG emission spectrum overlaid, while the lower panel displays the corresponding radiation-induced absorption coefficient $\mu_{\mathrm{ind}}$ as a function of the wavelength.

\rev{The transmission loss is concentrated in the 540 to 600~nm region, which overlaps the GAGG emission band.}

After irradiation, the total absorption coefficient $\mu_{\mathrm{after}}$ includes both intrinsic absorption and additional contributions from radiation-induced defects ($\mu_{\mathrm{ind}}$), such as color centers. These defects introduce new absorption bands within the material, increasing the overall optical absorption. Consequently, the optical transmission decreases (as seen in Fig.~\ref{fig:gagg_irrad}), resulting in increased $\mu$ and reduced attenuation length.

An analysis of the radiation-induced attenuation length is presented in Section~\ref{sec:att_irrad}, where $\mu_{\mathrm{ind}}$ is quantitatively determined from the transmission measurements described in Section~\ref{sec:rad_meth}.

For comparison, non-ultrafast GAGG ($\tau_{\mathrm{d,eff}} \simeq 44$~ns) under similar 24~GeV proton irradiation shows an induced absorption of $\sim$3.6~m$^{-1}$ at the emission peak for a fluence of $3.1\cdot10^{15}$~protons$\cdot$cm$^{-2}$~\cite{Alenkov2019}. The corresponding value for our sample is $\mu_{\mathrm{ind}}(535~\mathrm{nm}) \simeq 7.6$~m$^{-1}$ at a fluence of $4.0\cdot10^{15}$~protons$\cdot$cm$^{-2}$, that is, about a factor of two higher than in~\cite{Alenkov2019} for a fluence about 30\% larger. Mg co-doping has moreover been found to reduce the induced absorption under $\gamma$-irradiation, 0.6 against 1.5~m$^{-1}$ at 120~kGy~\cite{LUCCHINI2016176}, so the presence of Mg is not in itself detrimental to radiation tolerance. 
\rev{This assessment is based on the optical transmission of a single bulk sample, measured after a 3-month cool-down period.
 }

\subsection{Photoluminescence and radioluminescence spectra}

Regarding the PL spectra, the maximum luminescence intensity was observed at $\lambda$ = 540 nm for an excitation at $\lambda$ = 450 nm.

\begin{figure}[!ht]
    \centering
    \includegraphics[width = 1.\linewidth]{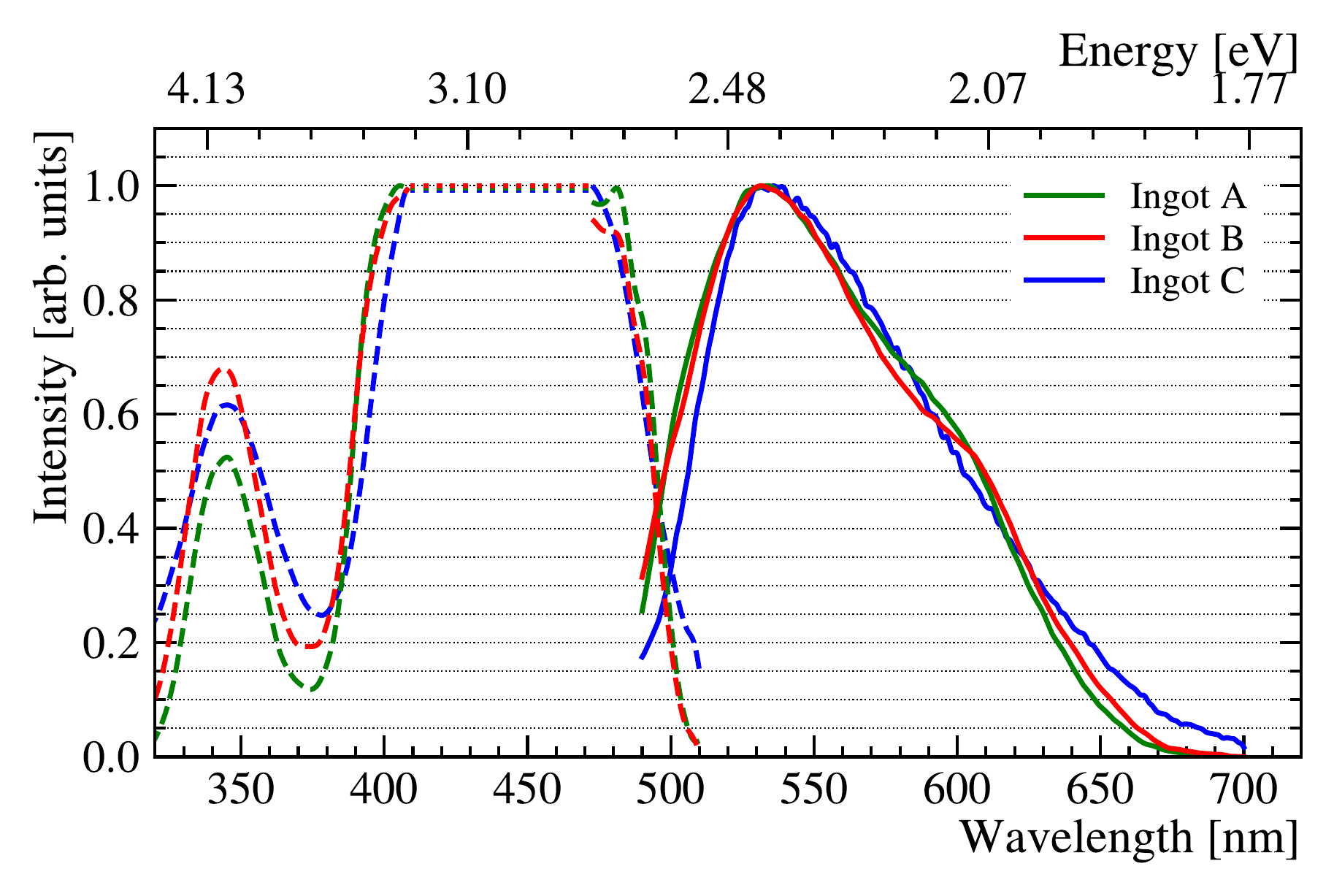}
    \caption{Emission and excitation spectra of the various ingots. The x-axis is presented in terms of wavelength (nm) and energy (eV). \textit{Dashed lines are representing the excitation spectra while solid lines are for emission spectra.}}
    \label{fig:emission}
\end{figure}

In Fig.~\ref{fig:emission}, we have both the emission and excitation spectra of ingots A, B and C.

We see that both the ingots B and A have similar emission and excitation spectra while the ones of ingot C are slightly shifted towards higher wavelength.

\begin{figure}[!ht]
    \centering
    \includegraphics[width = 1.\linewidth]{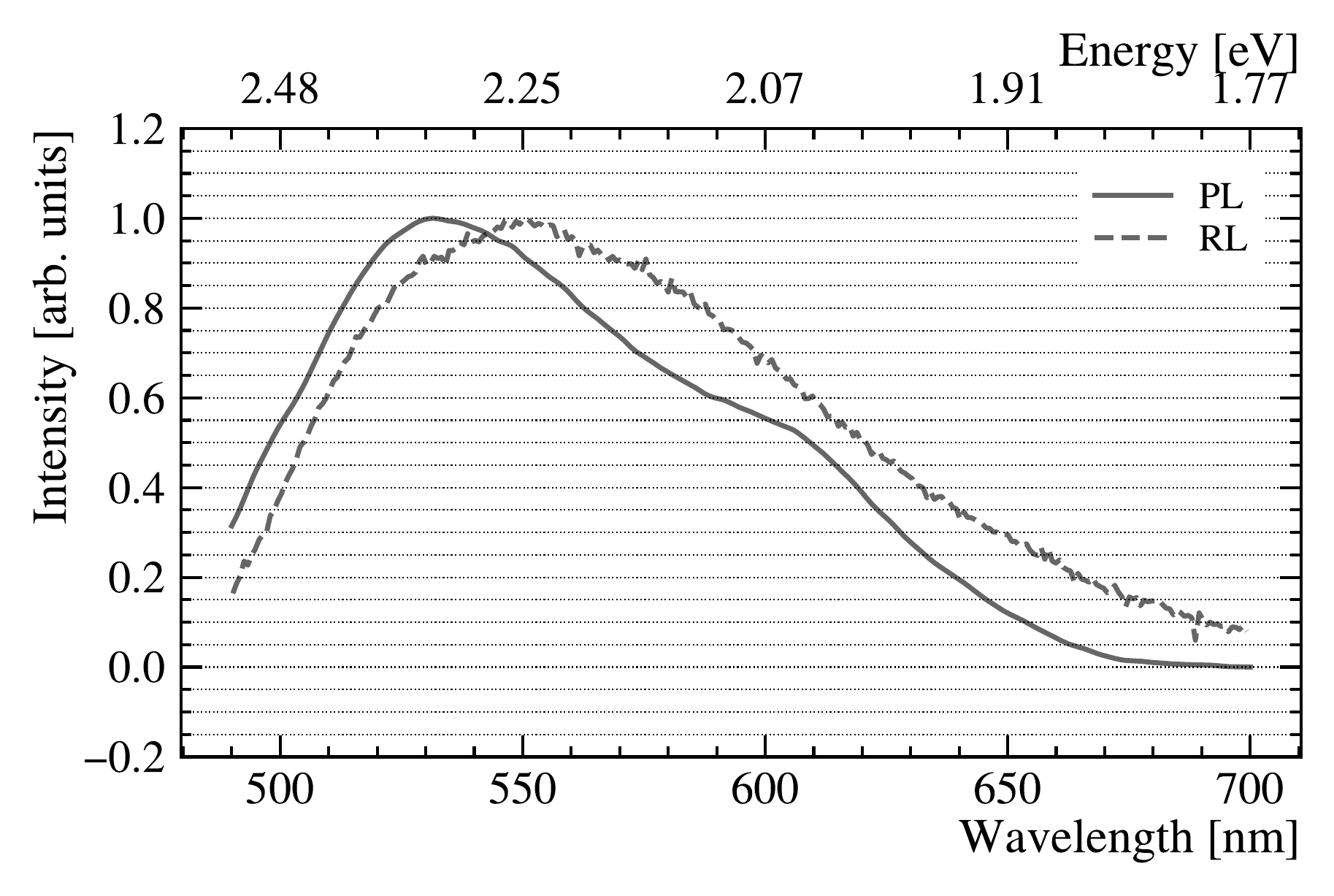}
    \caption{Photoluminescence and Radioluminescence spectra of the ingot A.}
    \label{fig:rl_vs_pl}
\end{figure}

We observe a redshift in the RL spectra compared to PL spectra as seen in Fig.~\ref{fig:rl_vs_pl}. This shift arises from the different measurement configurations: RL measured in transmission mode is subject to self-absorption due to the high dopant concentration, whereas PL measured in reflection mode minimizes this effect.

\subsection{Light attenuation}
We report here the attenuation length measurements of two crystal fibers with lengths of 5~cm and 10~cm, both extracted from ingot C. The light intensity ratio measured by the two spectrometers is normalized as left-to-right (L/R), as indicated in the figure labels.

\begin{figure}[!ht]
    \centering
    \includegraphics[width = 1.\linewidth]{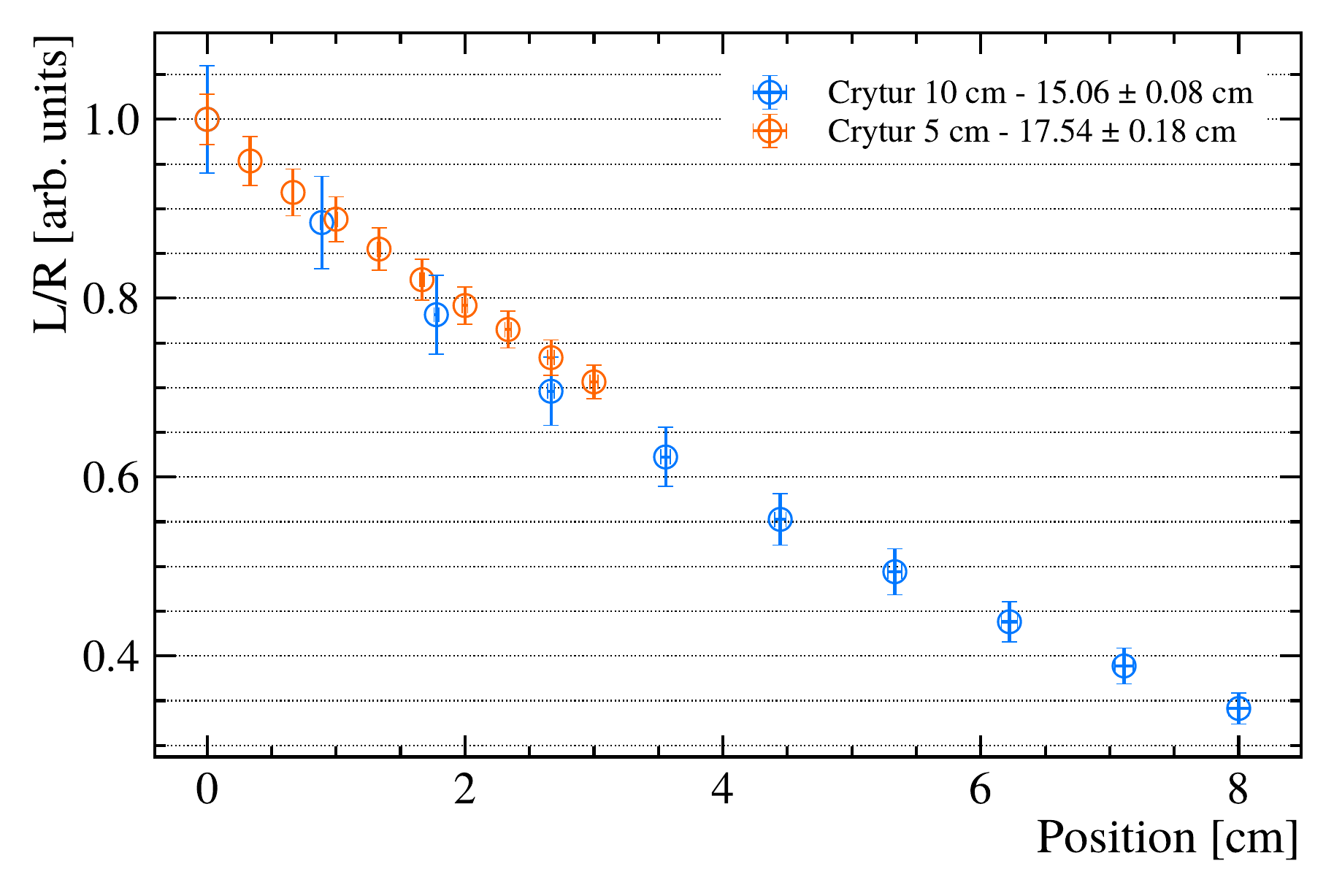}
    \caption{Attenuation length measurements for crystal fibers of 5 and 10 cm from ingot C. The data show the L/R ratio representing the normalized light intensity between left and right spectrometers.}
    \label{fig:att_fibers}
\end{figure}

Fig.~\ref{fig:att_fibers} presents the measured light intensity as a function of LED position along the crystal fiber, where position 0 corresponds to the first accessible point at the left end. Since 1~cm is excluded at each end, the scan spans $\sim$8~cm for the 10~cm crystal fiber and only $\sim$3~cm for the 5~cm one. The L/R ratio represents the left-to-right light intensity normalized at each position and computed integrating over the emission spectrum of GAGG. 
 
A small difference is observed between the 5~cm and 10~cm crystal fiber measurements, which is analysed below.

The effective attenuation length ($\lambda_{\mathrm{eff}}$) is then extracted fitting Eq.~\ref{eq:att_length}, providing an emission-weighted characteristic length that accounts for the full spectral distribution of the light.

\begin{equation}\label{eq:att_length}
    f(x) = a_0 \cdot \exp(- \frac{2 \cdot x} {\lambda_{\mathrm{eff}}})
\end{equation}

The obtained values for the 5 and 10 cm crystal fibers are respectively $17.54 \pm 0.18$~cm and $15.06 \pm 0.08$~cm.

Since $\lambda_{\mathrm{eff}}$ should ideally be an intrinsic bulk property, this difference reflects the measurement conditions and the distributed optical quality of the material rather than a fundamental one. We adopt the value from the 10~cm crystal fiber, $\lambda_{\mathrm{eff}} = 15.06 \pm 0.08$~cm, as the more reliable estimate: its usable scan spans $\sim$8~cm against only $\sim$3~cm for the 5~cm crystal fiber, giving a far better constrained exponential fit (Eq.~\ref{eq:att_length}), as reflected in its smaller uncertainty ($\pm0.08$ against $\pm0.18$~cm). Being longer, it also probes a longer light path and therefore samples the distributed bulk scattering and residual inclusions of these experimental ingots more representatively, consistent with the slightly shorter value it yields. Its $\sim$8~cm probed path moreover falls within the 5--10~cm operating length of an actual SpaCal crystal fiber, whereas the $\sim$3~cm probed by the 5~cm crystal fiber does not.

For comparison, non-ultrafast GAGG crystal fibers ($\tau_{\mathrm{d,eff}} \simeq 60$~ns) of the same geometry, measured with a similar technique, reach an attenuation length of $101.5$~cm~\cite{Martinazzoli2020}. The shorter value obtained here is a likely cost of the heavy Ce/Mg co-doping.

\subsection{Light yield}\label{sec:ly}
The results of the light yield (LY) and energy resolution measurements are reported in Table~\ref{tab:ly_GAGG}, assuming a relative uncertainty of 5\% for both quantities. The values were determined by fitting both the photopeak and the Compton edge of the acquired spectra for each sample. For comparison, the corresponding values for the commercial fast GAGG grade (GFAG, C\&A) are also included in the table.

\begin{figure}[!ht]
    \centering
    \includegraphics[width = \linewidth]{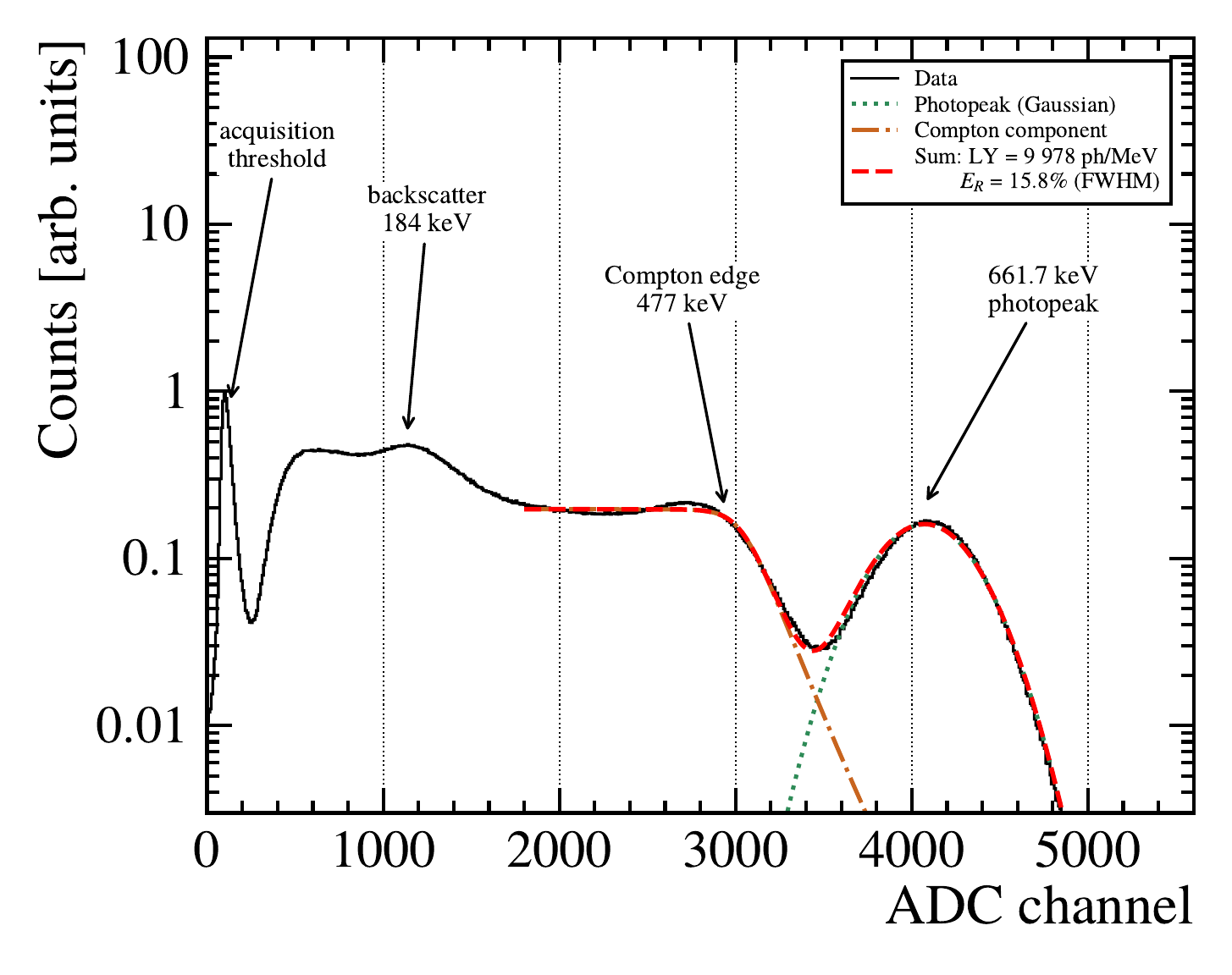}
    \caption{\rev{{$^{137}$Cs} spectrum of the {$1 \times 10$}~(\diameter)~mm$^3$ GAGG plate of ingot~C, measured on the light yield bench of Section~{\ref{sec:ly_meth}}. The 661.7~keV photopeak is fitted with a Gaussian (dotted), the Compton continuum with a smooth step function (dash-dotted), and their sum is shown in red (dashed). The backscatter peak (184~keV), the Compton edge (477~keV) and the acquisition threshold are indicated. The light yield and the energy resolution quoted in the legend are the values listed for this sample in Table~{\ref{tab:ly_GAGG}}.}}
    \label{fig:cs137_ingotC}
\end{figure}

\rev{A typical acquired spectrum is shown in Fig.~{\ref{fig:cs137_ingotC}} for the {$1 \times 10$}~(\diameter)~mm$^3$ plate of ingot~C, together with the fit from which both quantities are extracted.}

\rev{The acquired spectra are fitted with the function given in Eq.~{\ref{eq:cs137_fit}}, the sum of a Gaussian photopeak and a smooth step reproducing the Compton continuum.}

\begin{equation}\label{eq:cs137_fit}
    f(x) = A \cdot \exp \left( - \frac{(x - \mu)^2}{2 \sigma^2} \right) + \frac{a_0}{\sqrt{1 + \left( \frac{x}{a_1} \right)^{a_2}}}
\end{equation}

\rev{Here {$x$} is the ADC channel, while {$A$}, {$\mu$} and {$\sigma$} are the amplitude, the position and the width of the 661.7~keV photopeak, and {$a_0$}, {$a_1$} and {$a_2$} are the amplitude, the edge position and the sharpness of the step. The light yield is obtained from {$\mu$} through the single photo-electron response of the PMT (Section~{\ref{sec:ly_meth}}), and the energy resolution (FWHM) as {$E_R = 2.355 \, \sigma / \mu$}.}

The LY across the tested ingots ranges from 3\,207 ph/MeV (ingot B) to a maximum of 13\,273 ph/MeV (ingot D). A clear ``geometry effect'' is observed: for most ingots the plate-shaped samples provide the highest light collection efficiency.

The absolute LY (a few thousand up to $\sim$13\,000~ph/MeV) is markedly lower than the $\sim$50\,000~ph/MeV of non-ultrafast GAGG:Ce. This is the expected trade-off of the acceleration: the heavy Mg co-doping and closely spaced Ce--Mg pairs introduce non-radiative quenching that shortens the decay time to a few nanoseconds at the cost of photon yield.

\begin{table*}[ht]
    \centering
    \caption{Light yield (LY) and energy resolution ($E_R$) FWHM values for various GAGG samples measured with a $^{137}$Cs source.}
    \label{tab:ly_GAGG}
    \begin{tabular}{l|cc|cc|cc}
    \hline
    \hline
    & \multicolumn{2}{c}{\textbf{Plates ($1 \times 10$ (\diameter))}} & \multicolumn{2}{c}{\textbf{Pixels ($2\times2\times3$)}} & \multicolumn{2}{c}{\textbf{$2\times2\times10$}}\\
    Ingot & LY [ph$\cdot$MeV$^{-1}$] & $E_R$ [\%] & LY [ph$\cdot$MeV$^{-1}$] & $E_R$ [\%] & LY [ph$\cdot$MeV$^{-1}$] & $E_R$ [\%]\\
    \hline
    A  & 7\,426  & 21.2 & 7\,441 & 19.1 & 5\,157 & 22.5 \\
    B  & 3\,980  & 22.4 & 5\,075 & 22.0 & 3\,207 & 22.0\\
    C  & 9\,978  & 15.8 & 9\,758 & 16.1 & 6\,905 & 20.9\\
    D & 13\,273 & 18.8 & $\emptyset$  & $\emptyset$ & $\emptyset$ & $\emptyset$\\
    \hline
    GFAG C\&A \cite{Martinazzoli2021} & $\emptyset$ & $\emptyset$ & 32\,140  & 11.3 & $\emptyset$ & $\emptyset$\\
    \hline
    \hline
    \end{tabular}
\end{table*}

Variations among the geometries are primarily attributed to self-absorption effects arising from the different optical depths associated with varying crystal cross-sections. All geometries were optically coupled to the PMT photocathode via their largest face. 

The measured energy resolution values (expressed as the Full Width at Half Maximum (FWHM)) varied across the different ingots: 21.2\% for ingot A, 22.4\% for ingot B, 15.8\% for ingot C, and 18.8\% for ingot D. Notably, ingot C exhibited the superior performance in terms of energy resolution among the characterized samples.

\subsection{Scintillation kinetics}\label{sec:emiss}

\begin{figure}[ht]
    \centering
    \includegraphics[width = .9\linewidth]{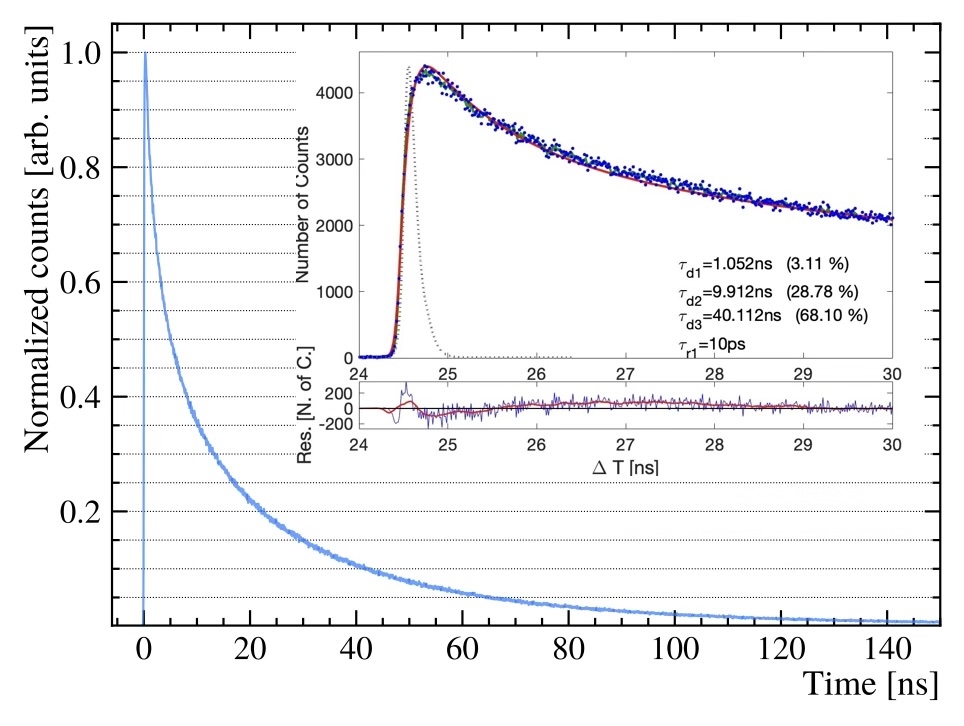}
    \caption{Decay time spectrum of a GAGG plate from ingot C. Measured scintillation decay time spectrum with a zoomed inset showing the fit and corresponding fit parameters.}
    \label{fig:scinti_plot}
\end{figure}

\begin{figure}[!ht]
    \centering
    \includegraphics[width = \linewidth]{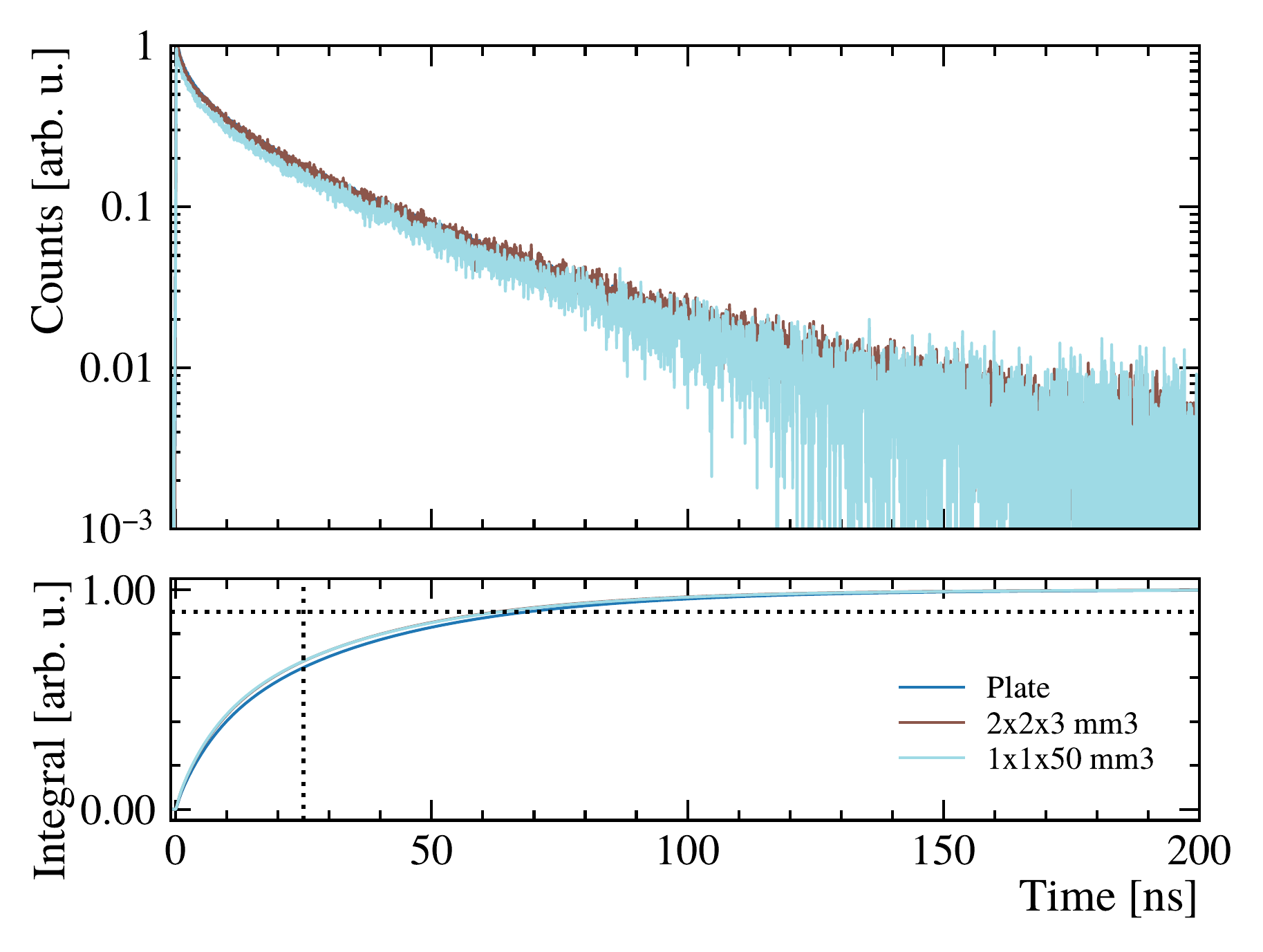}
    \caption{\textit{Top:} Superposition of GAGG samples pulses distributions measured in the X-ray bench for various geometries of the ingot C, log-scale for the $y$ axis. \textit{Bottom:} integral value as a function of time in ns.}
    \label{fig:super_GAGG_plate}  
\end{figure}
 
\begin{table*}[ht]
    \centering
        \caption{Decay time values of GAGG samples measured under X-ray excitation. The uncertainty is 25 ps and 5\% for the rise and decay times. The values for GFAG (C\&A) are taken from~\cite{Martinazzoli2021}, and its $\tau_{1/e}$ was computed from the reported decay components.}
    \label{tab:decay_time_GAGG}
    \setlength{\tabcolsep}{4pt}
    \small
    \begin{tabular*}{\textwidth}{@{\extracolsep{\fill}}lcccccccccc}
    \hline
    \hline
       Ingot - Geometry & $\tau_{\mathrm{r}}$ [ps] & $\tau_{\mathrm{d},1}$ [ns] & $\rho_1$ [\%] & $\tau_{\mathrm{d},2}$ [ns] & $\rho_2$ [\%] & $\tau_{\mathrm{d},3}$ [ns] & $\rho_3$ [\%] & $\tau_{\mathrm{d,eff}}$ [ns]  & $\tau_{\mathrm{d,average}}$ [ns] & $\tau_{1/e}$ [ns]\\
    \hline

A - Plate & 19 & 1.0 & 4.6 & 8.7 & 35 & 35 & 61 & 9.8 & 24.5 & 6.9\\
A - Pixel & 11 & 0.6 & 2.8 & 6.7 & 29 & 32 & 68 & 8.8 & 24.0 & 6.5 \\
B - Plate & 17 & 0.7 & 6.3 & 5.3 & 38 & 24 & 56 & 5.5 & 15.4 & 4.0\\
B - Pixel & 15 & 0.7 & 3.7 & 6.3 & 32 & 29 & 65 & 8.0 & 20.5 & 5.9\\ 
C - Plate & 10 & 1.0 & 3.1 & 9.9 & 29 & 40 & 68 & 13.2 & 30.2 & 9.7\\
C - Pixel & 14 & 0.7 & 2.3 & 8.0 & 24 & 36 & 73 & 12.0 & 28.5 & 8.9\\
D - Pixel & 12 & 1.1 & 1.8 & 12.8 & 26 & 53 & 72 & 19.5 & 41.5 & 14.8\\
\hline
GFAG C\&A - Pixel & 63 & 41 & 65 & 172 & 35 & $\emptyset$ & $\emptyset$ & 56.0 & 86.8 & 47.4\\
    \hline
    \hline
    \end{tabular*}
\end{table*}

Decay time measurements were performed on all GAGG samples using the TCSPC setup. A representative decay spectrum fit is shown in Fig.~\ref{fig:scinti_plot}, while Fig.~\ref{fig:super_GAGG_plate} presents spectra for the full sample set.

All measured decay spectra were properly fitted using Eq.~\ref{eq:scinti_eq_bso}, which models the scintillation response with one rise time component and three exponential decay components. 

The fitted parameters reveal that all samples exhibit effective decay times ($\tau_{\mathrm{d,eff}}$) below 20~ns, with many under 10~ns. Across the sample set, the dominant decay component, carrying the majority of the scintillation intensity, is the slowest component ($\tau_{\mathrm{d},3}$), which ranges from 24 to 53~ns, while the fastest component consistently stays around or below 1 ns.
Table~\ref{tab:decay_time_GAGG} summarizes all extracted parameters, including $\tau_{\mathrm{d,eff}}$, $\tau_{\mathrm{d,average}}$, and $\tau_{1/e}$, with an estimated uncertainty of approximately 5\% on all reported decay time values. For comparison, the corresponding values for the commercial fast GAGG grade (GFAG, C\&A) are also included in the table.

\subsection{Time resolution}\label{sec:ctr}
\subsubsection{With gamma}

\begin{figure}[ht]
    \centering
    \includegraphics[width = \linewidth]{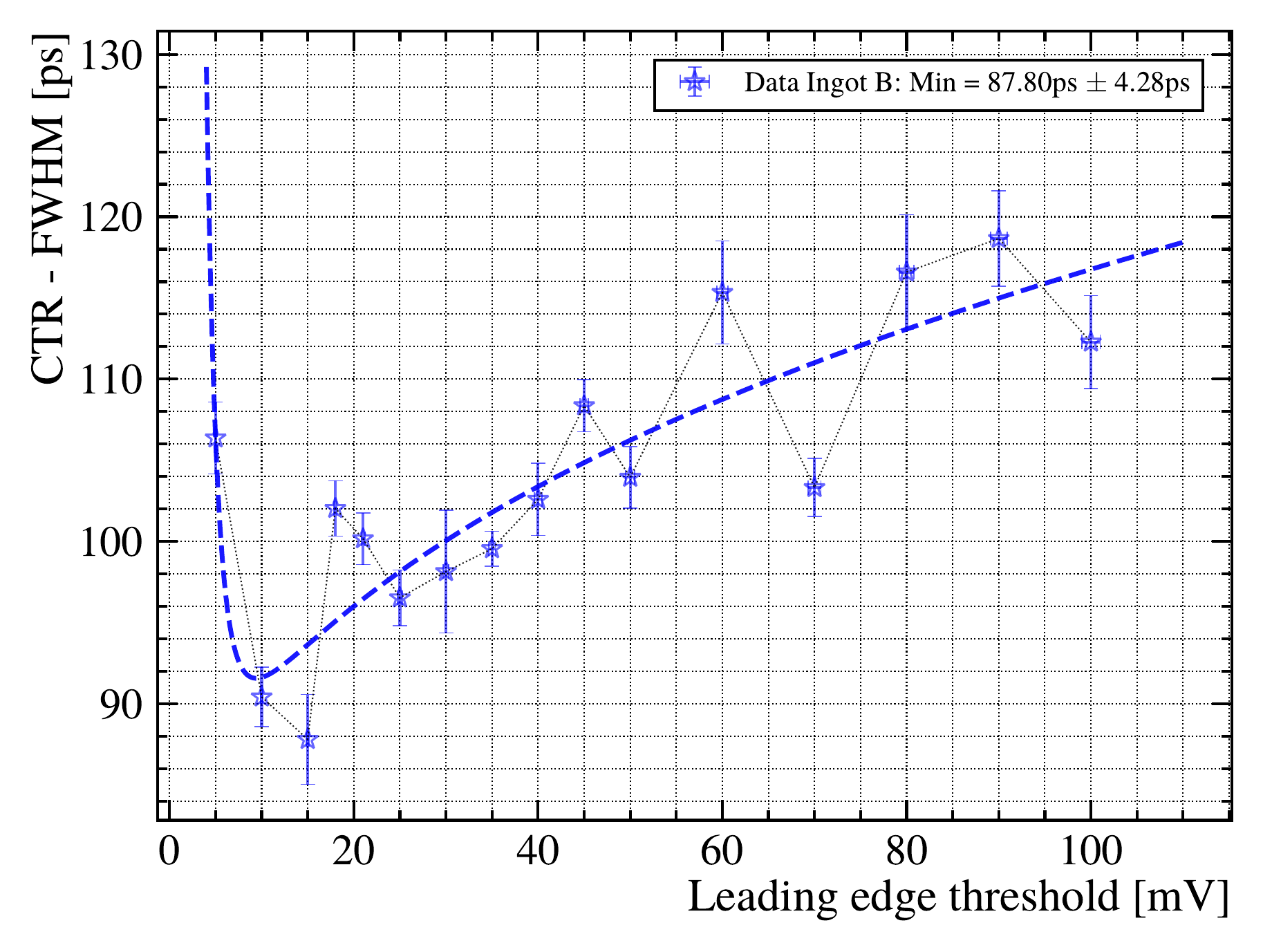}
    \caption{Scan of the leading edge threshold for the GAGG from ingot B measured with NUV-HD-MT SiPM using $^{22}$Na source.}
    \label{fig:compare_cryt}
\end{figure}

For the CTR measurements, two pixels from the same ingot were placed face-to-face in coincidence.
The complete set of measured CTR values is reported in Table~\ref{tab:ctr_GAGG}. Fig.~\ref{fig:compare_cryt} shows the time resolution as a function of the leading edge discrimination threshold for a sample from ingot B.

We also measured the CTR of a GFAG sample from C\&A, representing ``commercial-grade'' GAGG, to benchmark our accelerated GAGG compositions. Its light yield and decay time, measured on the same benches, are reported in Tables~\ref{tab:ly_GAGG} and~\ref{tab:decay_time_GAGG}.

 \begin{table}[ht]
\centering
\caption{Results of CTR (FWHM) measured with HF electronics using a $^{22}$Na source.}
\label{tab:ctr_GAGG}
\begin{tabular}{lc}
\hline \hline Sample - 2$\times$2$\times$3 mm$^3$ & CTR FWHM [ps]\\
\hline
GAGG Crytur B  & $88 \pm 4$ \\
GAGG Crytur C  & $91 \pm 4$ \\
GFAG C\&A & $93 \pm 4$ \\
LYSO:Ce,Ca & $66 \pm 2$ \\
\hline \hline
\end{tabular}
\end{table}

\subsubsection{With 120 GeV pions}

As a reference benchmark, we measured LYSO:Ce,Ca crystals from Taiwan Applied Crystals, achieving a CTR FWHM of 66~ps with gamma sources under the previous setup conditions. LYSO:Ce represents the current state-of-the-art in fast timing scintillators \cite{Gundacker_2016}, and has been selected for the Barrel Timing Layer (BTL) of the CMS experiment at CERN \cite{Addesa_2024} due to its outstanding timing performance, high light yield, and radiation tolerance.

Our study indicates time resolutions ($\sigma$) of $13.1 \pm 0.7$~ps for accelerated GAGG and $12.6 \pm 0.8$~ps for LYSO:Ce,Ca. Following the usual convention, time resolutions measured with high-energy particles are quoted as Gaussian $\sigma$, whereas the CTR values of Section~\ref{sec:ctr} are quoted as FWHM. This result confirms that accelerated GAGG achieves timing capabilities equivalent to the LYSO:Ce,Ca benchmark under 120 GeV pions, establishing it as a competitive alternative for precision timing applications in particle physics detectors.

\begin{figure}[ht]
    \centering
    \includegraphics[width = \linewidth]{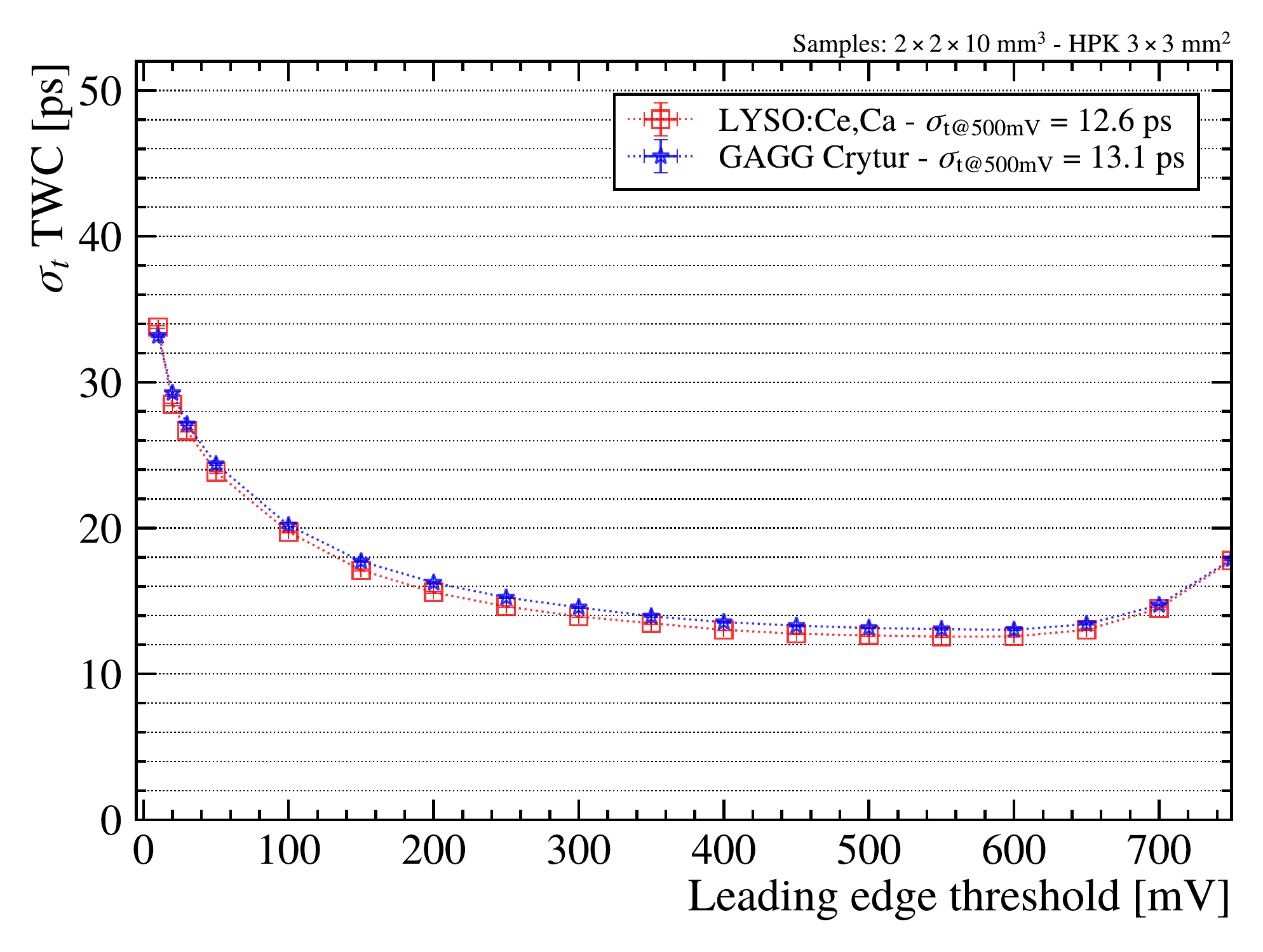}
    \caption{Scan of the leading edge threshold for a $2\times2\times10$ mm$^3$ GAGG from ingot C (blue) and LYSO:Ce,Ca (red) measured with HPK S13360-3050PE under a 120 GeV pion beam.}
    \label{fig:tres_mips}
\end{figure}

These values are the minima of the leading-edge threshold scans in Fig.~\ref{fig:tres_mips}, which plot the time resolution of GAGG (ingot C) and LYSO:Ce,Ca against the discrimination threshold. The GAGG curve closely tracks that of LYSO:Ce,Ca over the full scan, which is the basis for the equivalent-performance claim.

We note that the optimal detection threshold differs substantially between measurement configurations: $\approx 10$ mV for CTR measurements and up to $500$ mV for 120 GeV pion detection, reflecting the distinct particle types and energy depositions in each scenario.

\section{Discussion}\label{sec:disc}

\subsection{Optical properties}

\subsubsection{Emission and absorption}

\begin{figure}[ht]
    \centering
    \includegraphics[width = \linewidth]{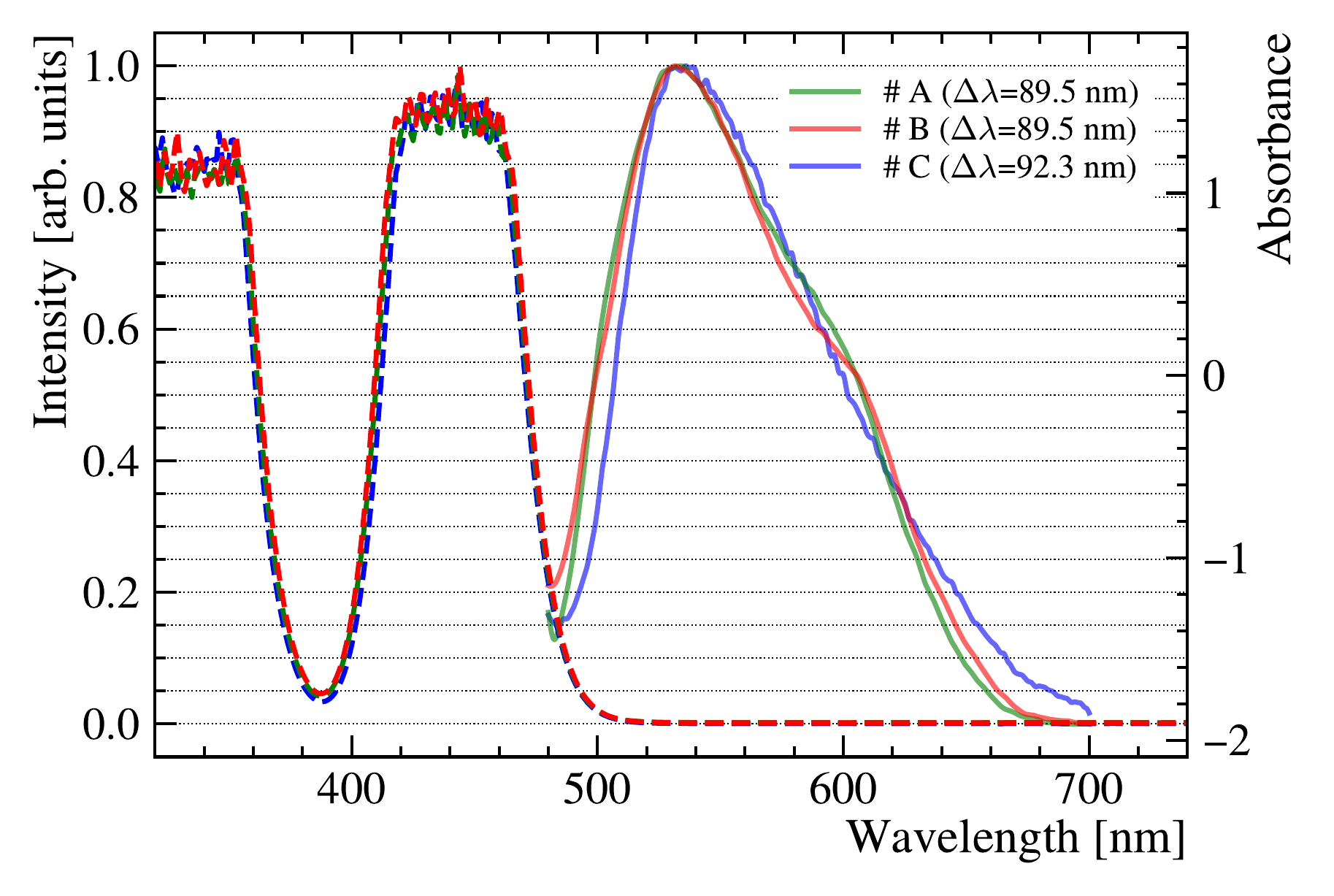}
    \caption{Absorption and Emission spectra of the ingots A, B and C.}
    \label{fig:abs_vs_em}
\end{figure}

The Stokes shift, expressed as $\Delta \lambda$ in Fig.~\ref{fig:abs_vs_em}, is large enough to avoid strong self-absorption, and heavy co-doping does not appear to alter this property. A residual overlap nevertheless remains in the blue tail of the emission: it shows up as the redshift of the RL spectrum with respect to the PL one (Fig.~\ref{fig:rl_vs_pl}), and in the geometry dependence of the light yield (Section~\ref{sec:ly}).

While a qualitative comparison provides initial insights, a quantitative analysis of the Ce-related absorption band is required in order to assess potential differences in the Ce charge state. In particular, the width $w$ of the broad absorption feature extending approximately from 400~nm to 500~nm is of interest. The width of this band is proportional to the Ce$^{3+}$ ion concentration, providing an indirect but sensitive probe of the activator content in the crystal.

\begin{figure}[ht]
    \centering
    \includegraphics[width = \linewidth]{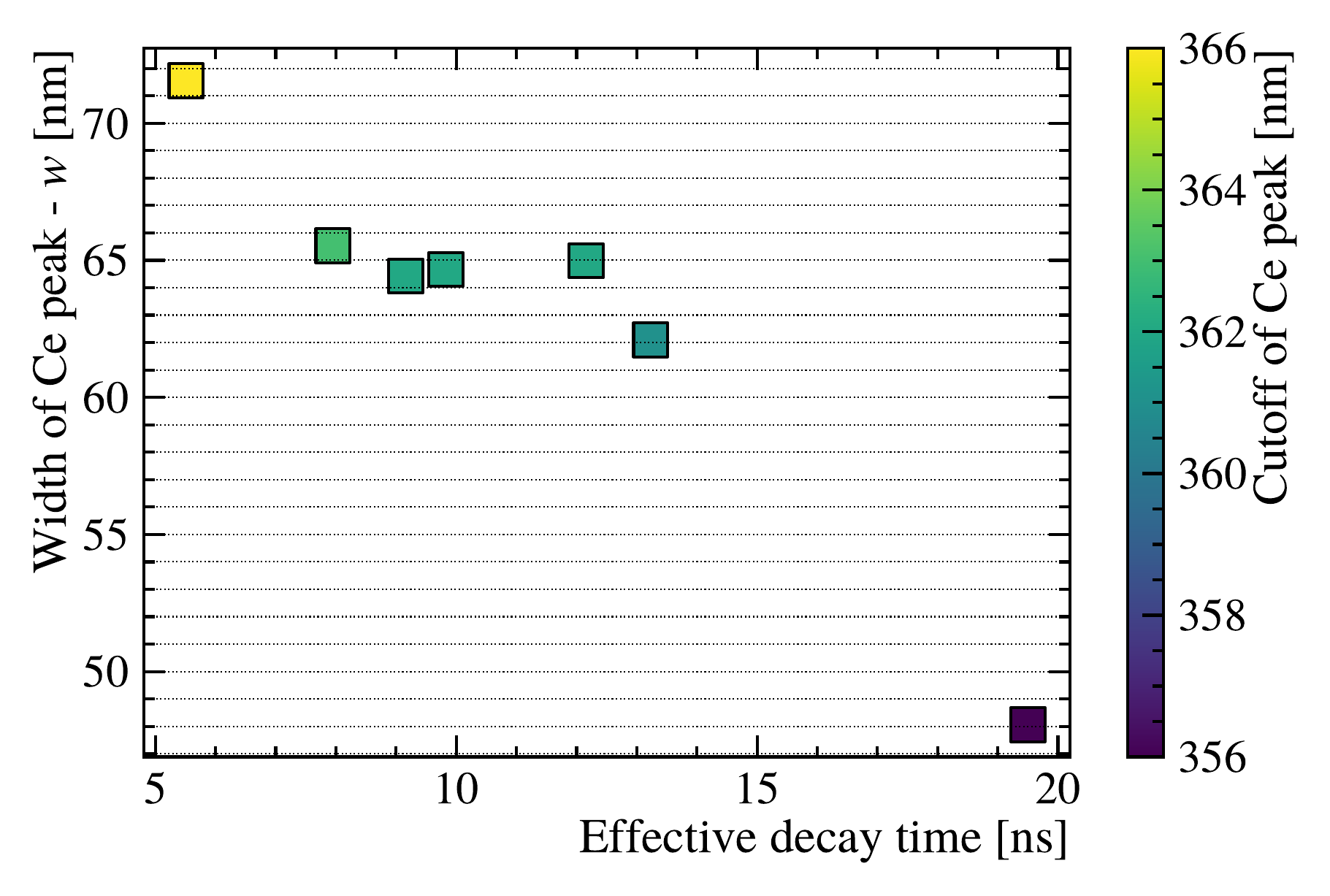}
    \caption{Ce peak width (w) as a function of effective decay time and cutoff for GAGG plates.}
    \label{fig:gagg_heatmap_decay_ly}
\end{figure}

Beyond the emission peak width, the absorption spectrum shows a cutoff wavelength at approximately 350 nm (Fig.~\ref{fig:gagg_absorbance}) which is mostly determined by the charge transfer absorption of Ce$^{4+}$ as the $4f\mapsto5d_{2}$ absorption transition of Ce$^{3+}$ peaking within 330--340 nm is completely hindered. As illustrated in Fig.~\ref{fig:gagg_heatmap_decay_ly}, both parameters: width $w$ and cutoff wavelength, show a strong correlation with the effective decay time.

\subsubsection{Attenuation and irradiation}\label{sec:att_irrad}

\begin{figure}[ht]
    \centering
    \includegraphics[width = \linewidth]{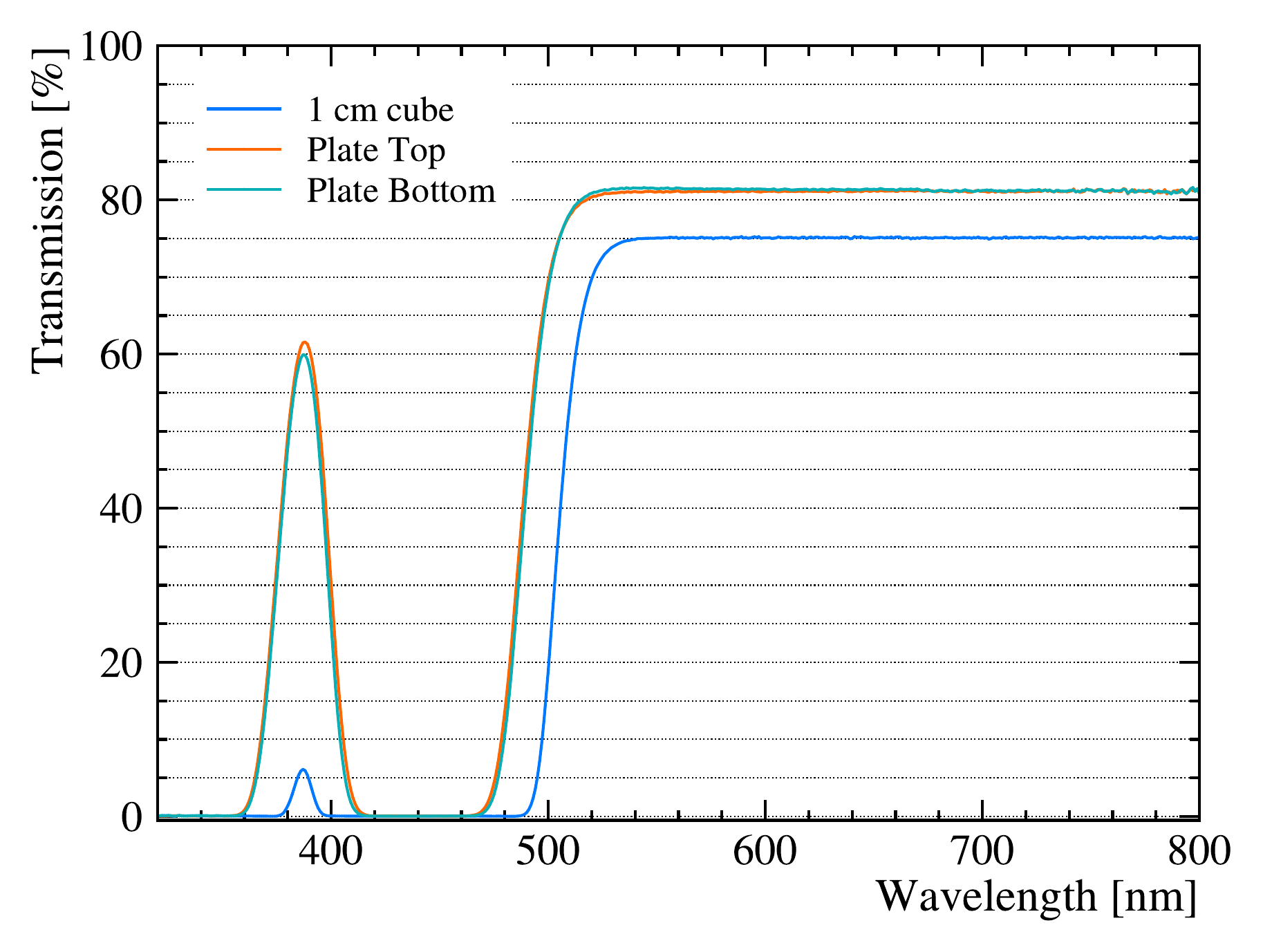}
    \caption{Transmission of GAGG samples from ingot C.}
    \label{fig:transmission_ingot99}
\end{figure}

To ensure the consistency of the optical properties across different geometries, two samples with different dimensions: a cube and a plate, were prepared from ingot~C. The optical path lengths of the cube and the plate are $x_2 = 10~\mathrm{mm}$ and $x_1 = 1~\mathrm{mm}$, respectively. As shown in Fig.~\ref{fig:transmission_ingot99}, the measured transmittance above $550~\mathrm{nm}$ is approximately $T_2 = 75.0\%$ for the cube and $T_1 = 80.6\%$ for the plate.

Assuming identical surface conditions for both samples, the bulk attenuation coefficient can be estimated using the differential Beer-Lambert relation as given in Eq.~\ref{eq:abs_approx}.
\begin{equation}
\label{eq:abs_approx}
\mu = - \frac{\ln(T_2 / T_1)}{x_2 - x_1},
\end{equation}
which removes surface reflection losses by taking the ratio of transmissions for samples of identical composition but different thicknesses, following the method described in~\cite{benaglia2026characterizationopticalfiltershigh}.

This yields a bulk attenuation length of $\lambda = \mu^{-1} = 12.50 \pm 0.45~\mathrm{cm}$ at wavelengths above $550~\mathrm{nm}$. This is of the same order as the $15.06~\mathrm{cm}$ obtained independently from the crystal fiber measurements (Fig.~\ref{fig:att_fibers}). The two are not expected to coincide exactly: the differential estimate is a bulk value restricted to wavelengths above 550~nm, whereas the crystal fiber value is weighted over the full emission spectrum, as discussed below.

As a consistency check, Fresnel reflections were also considered using the refractive index of GAGG ($n \approx 1.90$ at $550~\mathrm{nm}$). 
The reflection coefficient at normal incidence is
\begin{equation}
R = \left( \frac{n - 1}{n + 1} \right)^2 \approx 0.096 .
\end{equation}
Including multiple internal reflections, the transmittance of a slab is given by
\begin{equation}
T = \frac{(1-R)^2 \cdot  e^{-\mu x}}{1 - R^2 \cdot e^{-2\mu x}}.
\end{equation}
Applying this expression to the measured transmittances of the plate and cube yields an attenuation length of approximately $12.0~\mathrm{cm}$, consistent with the value obtained from the differential method.

These results indicate that the bulk optical attenuation is comparable across the investigated geometries, supporting the use of the crystal fiber derived attenuation length as a reference for evaluating irradiation-induced changes.

However, scintillation light is emitted over a broad wavelength range, and its transport in the detector depends on the wavelength-dependent absorption of the material. To obtain values representative of the actual light propagation, including self-absorption effects, the absorption coefficients given hereafter were integrated over the GAGG emission spectrum.

The \textit{pre}-irradiation effective absorption coefficient for the crystal fiber, was determined to be $\mu_{\mathrm{before}} = 0.066~\mathrm{cm}^{-1}$ (corresponding to $\lambda_{\mathrm{before}} = 15.06~\mathrm{cm}$). $\lambda_{\mathrm{before}}$ is longer than the peak intrinsic length (i.e. the bulk attenuation length evaluated at the peak emission wavelength, $\sim$535~nm) because the weighting includes longer wavelengths where GAGG is more transparent.

Furthermore, while the differential bulk calculation (Eq.~\ref{eq:abs_approx}) does not take into account surface effects, the crystal fiber measurement (Fig.~\ref{fig:att_fibers}) accounts for Fresnel losses and light-trapping efficiency.

The radiation-induced absorption $\mu_{\mathrm{ind}}$ was quantified by comparing transmission spectra of the bulk cube before and after $1~\mathrm{MGy}$ proton irradiation (Eq.~\ref{eq:ind_abs}). Like $\mu_{\mathrm{before}}$, the value entering the calculation below is integrated over the GAGG emission spectrum, $\mu_{\mathrm{ind}} \simeq 0.93$~m$^{-1}$. It is much smaller than the $\mu_{\mathrm{ind}}(535~\mathrm{nm}) \simeq 7.6$~m$^{-1}$ quoted in Section~\ref{sec:rad_res} because the induced absorption band is narrow and vanishes above $\sim$600~nm (Fig.~\ref{fig:gagg_irrad}), whereas a large part of the GAGG emission is emitted at those longer wavelengths.

By applying this material degradation factor to the crystal fiber's initial performance, the \textit{post}-irradiation attenuation length ($\lambda_{\mathrm{after}}$) is:

\begin{equation}
\lambda_{\mathrm{after}} =
(\mu_{\mathrm{before}} + \mu_{\mathrm{ind}})^{-1}
\approx 13.28~\mathrm{cm}
\end{equation}

This corresponds to a reduction in attenuation length from $15.06~\mathrm{cm}$ to $13.28~\mathrm{cm}$, that is, a loss of about 12\%. Despite the 1~MGy dose, the attenuation length remains comparable to the crystal fiber lengths of 5 to 10~cm used in compact calorimeter applications such as SpaCal. For reference, non-ultrafast GAGG crystal fibers lost a larger fraction of their attenuation length under a comparable 1~MGy proton dose, from 101.5 to 33.6~cm, about 67\%~\cite{Martinazzoli2020}, although that material starts from a considerably higher initial transparency.

\rev{This comparison concerns the transport of the scintillation light only, as the irradiated cube was characterized on the transmission bench.}

This analysis assumes uniform bulk properties and neglects possible wavelength-dependent surface effects or scattering contributions that may not fully cancel in the cube--plate comparison. Furthermore, the extrapolation from bulk irradiation measurements to crystal fiber performance assumes that the radiation-induced absorption $\mu_{\mathrm{ind}}$ is homogeneous and identical in both geometries. In practice, variations in defect formation and surface quality may lead to deviations from the predicted attenuation length $\lambda_{\mathrm{after}}$.
Nonetheless, these potential discrepancies are mitigated by a comparison with theoretical transmission models; after accounting for Fresnel reflection losses ($(1-R)^2$), the resulting intrinsic attenuation values remain highly consistent with those derived from our experimental measurements.

\subsection{Light yield and scintillation kinetics}

Of particular importance for HEP applications is the longitudinal LY uniformity ($\Delta_z$~LY), which characterizes scintillation output variation along the crystal growth axis. This parameter critically affects detector response uniformity and energy resolution.

Ingot C demonstrated excellent longitudinal uniformity, with LY decreasing from 9\,978 to 9\,191~ph/MeV between head and tail over 10~cm, corresponding to a gradient of $\sim$0.8\%/cm. This is well within the less than $2\%$/cm required to preserve the energy resolution of a SpaCal-type sampling calorimeter~\cite{Roux2025}.
In contrast, ingot B showed poor uniformity (6\,939 to 3\,980~ph/MeV, $\sim$4.3\%/cm gradient) exceeding acceptable limits, and additionally exhibited excessive brittleness preventing fabrication of 10~cm samples. This ingot was therefore excluded from further study. Ingot A exhibited acceptable uniformity (7\,426 to 6\,787~ph/MeV, $\sim$0.9\%/cm) but lower absolute LY than ingot C, making the latter the most promising candidate for detector applications, explaining why this composition was chosen for crystal fiber production.

\begin{figure}[ht]
    \centering
    \includegraphics[width = \linewidth]{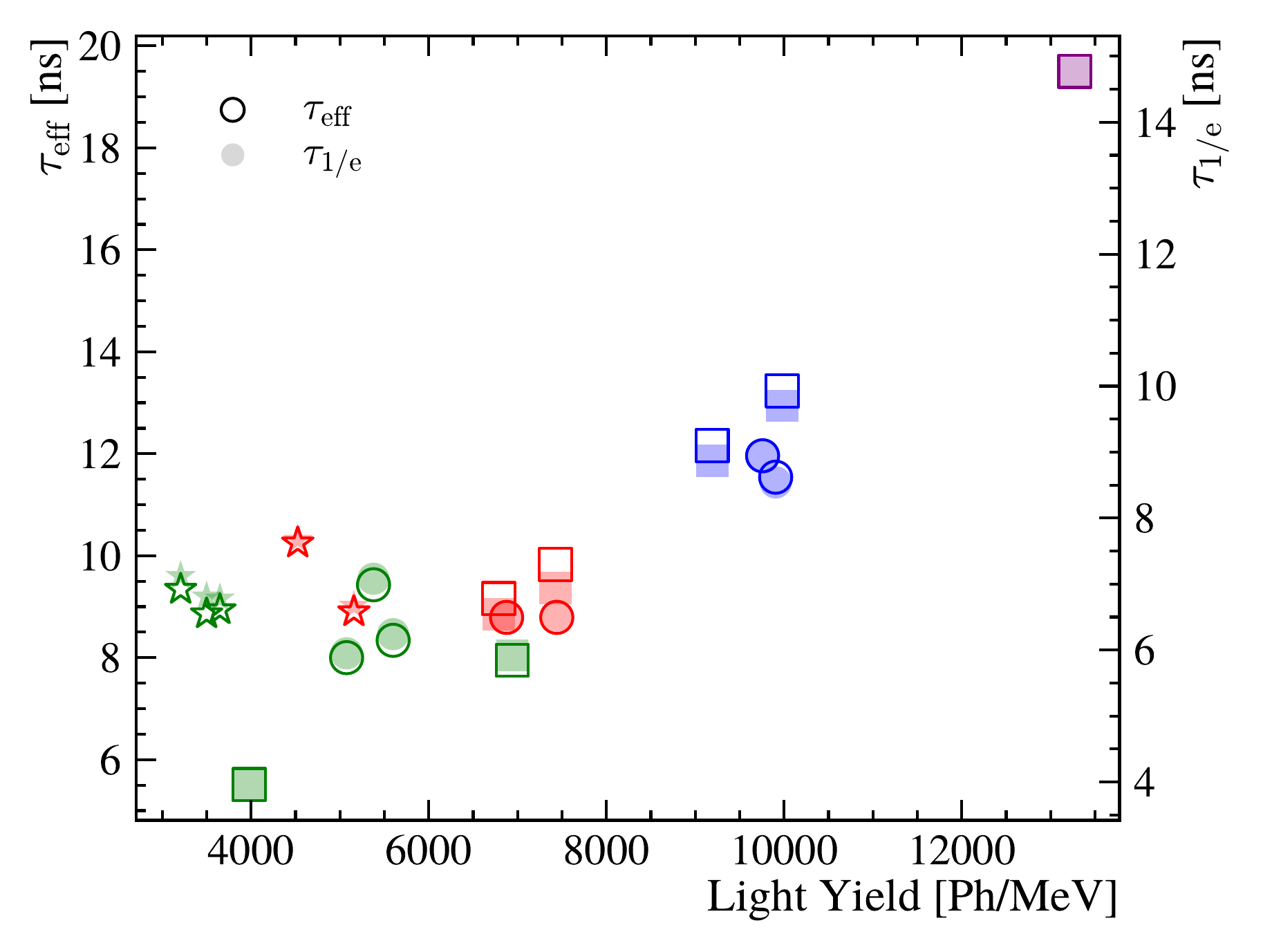}
    \caption{Effective decay time (empty markers) and $\tau_{1/e}$ (full markers) as a function of the light yield. \textit{Blue points: ingot C, green points: ingot B, red points: ingot A and the purple point: ingot D. Circles: $2\times2\times3$ mm$^3$, stars: $2\times2\times10$ mm$^3$ and squares represent plates.}}
    \label{fig:gagg_taueff_vs_ly}
\end{figure}

As illustrated in Fig.~\ref{fig:gagg_taueff_vs_ly}, both the $\tau_{\mathrm{d,eff}}$ and $\tau_{1/e}$ exhibit a clear linear relationship with the LY across the GAGG samples.

This linear dependence enables predictable tuning of GAGG properties for specific applications, for instance, timing-critical applications (TOF-PET, timing layers) benefit from faster decay times despite reduced photon statistics, while energy-resolution-driven applications favor maximized LY. 

Of particular relevance to HEP applications is the fraction of light emitted within 25 ns, the LHC bunch crossing interval, and the time at which 90\% of the total light has been emitted. Both quantities were extracted from the decay time spectra by computing the integral over the full gate (1 $\mu$s) and comparing it to the partial integrals at 25 ns and at the 90\% threshold, respectively.
Table~\ref{tab:time90_GAGG} gives the two additional parameters mentioned. The value of the integral at 25 ns is presented as a percentage relative to the full integral over the 1 $\mu$s gate. The time at which 90\% of the integral is reached is also provided.

\begin{figure}[ht]
    \centering
    \includegraphics[width = \linewidth]{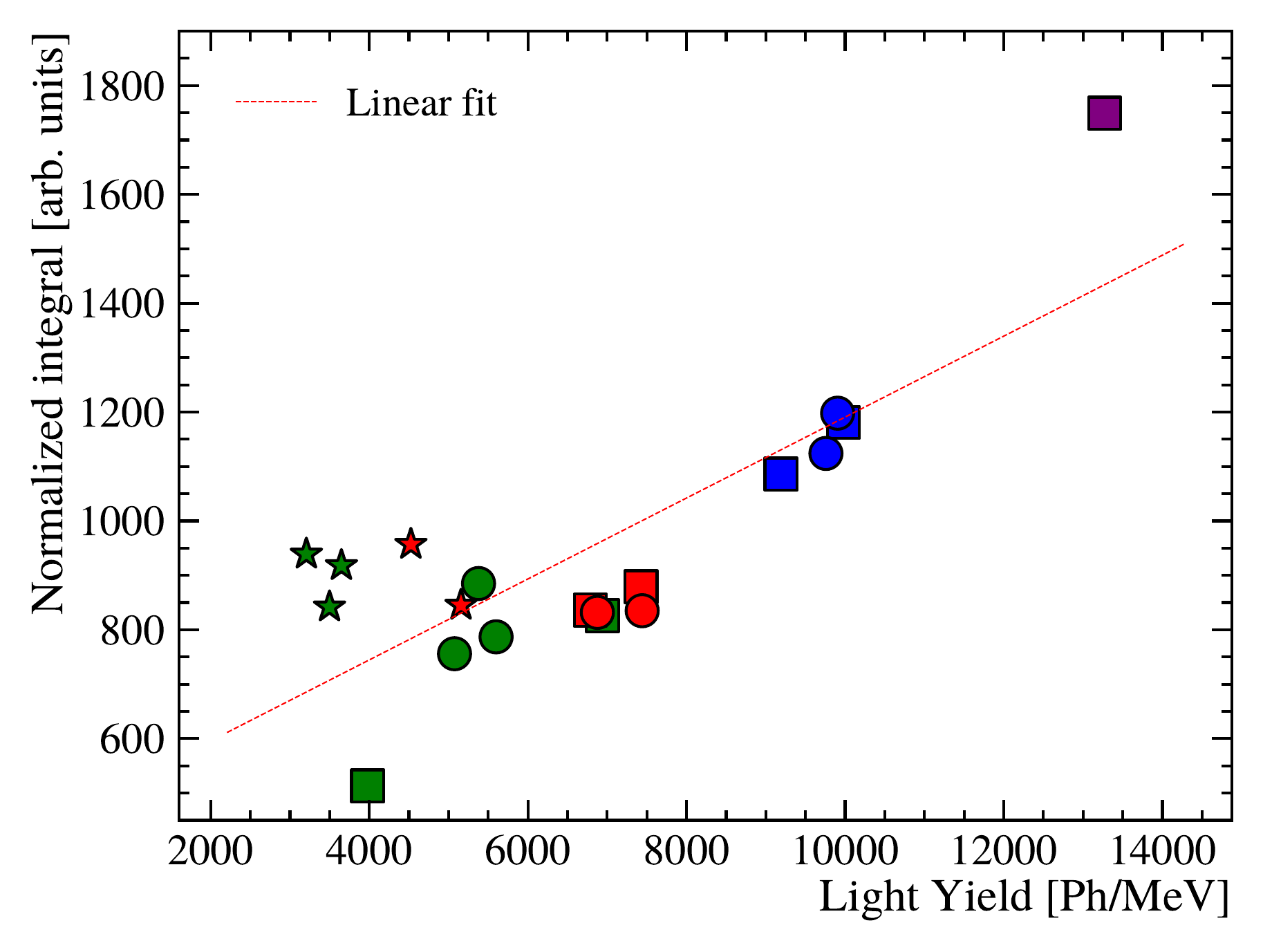}
    \caption{Integral of the decay time spectra (1 $\mu$s gate) with respective LY for different samples. \textit{Blue points: ingot C, green points: ingot B, red points: ingot A and the purple point: ingot D. Circles: $2\times2\times3$ mm$^3$, stars: $2\times2\times10$ mm$^3$ and squares represent plates.}}
    \label{fig:integral_vs_ly}
\end{figure}

Finally,Fig.~\ref{fig:integral_vs_ly} illustrates a good linear relationship between the integral value (over the 1 $\mu$s gate) and the LY of the samples, with a good correlation ($r^2 \simeq 0.70$).

\begin{table}[ht]
\centering
\caption{Percentage of the light emitted at 25 ns (I$_{25\mathrm{~ns}}$),  time for which 90\% of the light is emitted (T$_{90\%}$ ) and effective decay time of GAGG samples.}
\label{tab:time90_GAGG}
\begin{tabular}{cccc}
\hline \hline Ingot - Dimension & I$_{25\mathrm{~ns}}$ [\%] & T$_{90\%}$ [ns] & $\tau_{\mathrm{d,eff}}$ [ns]\\
\hline
A - Plate & 68.0 & 63.9 & 9.8 \\
A - Pixel & 67.7 & 62.2 & 8.8 \\
B - Plate & 80.0 & 38.2 & 5.5 \\
B - Pixel & 72.4 & 53.4 & 8.0 \\
C - Plate & 61.1 & 77.1 & 13.2 \\
C - Pixel & 62.1 & 72.2 &  12.0 \\
D - Pixel & 51.2 & 104.7 &  19.5\\
\hline \hline
\end{tabular}
\end{table}

While the effective decay time is well-established, previous studies have proposed additional metrics, such as $\tau_{1/e}$, along with others presented in Table~\ref{tab:time90_GAGG}. 

\subsection{Time resolution with gamma}

Our measurements confirm that the accelerated compositions achieve CTR (FWHM) values comparable to non-ultrafast GAGG despite lower absolute LY, demonstrating that sufficient photon statistics are maintained to preserve excellent timing performance. This result is particularly significant for HEP or ToF-PET applications, where an accelerated response enables improved performance at high count rates.

This can be explained by the fact that the CTR depends on both the timing characteristics and the LY according to Eq.~\ref{eq:ctr_prop}.

\begin{equation}\label{eq:ctr_prop}
    \mathrm{CTR} \propto \sqrt{\frac{\tau_{\mathrm{d,eff}} \cdot \tau_r}{\mathrm{LY}}}
\end{equation}

This relationship highlights a favorable trade-off for accelerated GAGG compositions: although heavy Ce and Mg doping reduces the LY, the significantly faster decay and rise times provide compensating benefits for timing performance. 
\section{Conclusions}

This work builds directly upon earlier R\&D efforts on GAGG \cite{Martinazzoli2021} and demonstrates a significant step forward in their timing performance and applicability to HEP. The crystals studied here are experimental and, as is common during the growth process, some variation in quality can occur from ingot to ingot. This is an inherent aspect of early-stage crystal development and can influence the measured optical and scintillation properties.
In this article we demonstrated that ultrafast GAGG compositions were repeatedly grown as long ingots (up to 10 cm), all consistently exhibiting effective decay times below 20 ns, reaching values as low as 5.5 ns. Crystal fibers cut from these ingots showed equally promising performance, confirming both the versatility of the growth process and the reproducibility of these ultrafast properties across different crystal geometries.
Despite this substantial acceleration of the scintillation process, light yields remain at the level of several thousand photons per MeV. Crucially, the accelerated composition retained most of its optical transmission up to 1~MGy under proton irradiation.
Timing measurements performed in the laboratory with gamma sources show time resolutions comparable to those of non-ultrafast commercial GAGG crystals. Moreover, measurements with 120 GeV pions indicate a timing performance that matches that of LYSO:Ce,Ca, the current state-of-the-art scintillator for fast-timing applications. These results confirm that this ultrafast GAGG combines fast scintillation kinetics with sufficient light output to achieve excellent time resolution.
\rev{In addition, the radiation-induced optical absorption measured after a 1~MGy proton dose is small, which is encouraging for use in high-radiation environments.} The attenuation length remains an area for further improvement before large-scale applications can be realized; its present value may be partly attributed to inclusions in the ingots, a consequence of both the growth variability mentioned above and the experimental nature of the ingots used in this work. Nevertheless, the results presented here establish this GAGG as a mature and competitive scintillator candidate for HEP applications such as calorimetry, \rev{combining ultrafast scintillation and optical radiation tolerance.}

\section*{CRediT authorship contribution statement}
\textbf{Louis Roux}: Formal analysis, Investigation, Methodology,
Software, Validation, Writing - original draft, Conceptualization, Writing - review \& editing. 
\textbf{Loris Martinazzoli}: Investigation, Methodology, Software, Validation, Conceptualization, Writing - review \& editing. 
\textbf{Julie Delenne}: Methodology, Software, Writing - review \& editing. 
\textbf{Philipp Roloff}: Resources, Supervision, Writing - review \& editing.
\textbf{Ondřej Zapadlík}: Growth of crystals, Writing - review \& editing.
\textbf{Jan Polak}: Growth of crystals, Writing - review \& editing.
\textbf{Jan Havlíček}: Growth of crystals, Writing - review \& editing.
\textbf{Silvia Sýkorová}: Growth of crystals, Writing - review \& editing.
\textbf{Martin Nikl}: Supervision, Writing - review \& editing.
\textbf{Pavel Boháček}: Supervision, Writing - review \& editing.
\textbf{Christophe Dujardin}: Resources, Supervision, Conceptualization,
Writing - review \& editing.
\textbf{Etiennette Auffray}: Resources, Supervision, Conceptualization,
Writing - review \& editing.

\section*{Declaration of competing interest}
The authors declare that they have no known competing financial interests or personal relationships that could have appeared to influence the work reported in this paper.

\section*{Acknowledgments}

This work was carried out in the frame of Crystal Clear Collaboration and has been supported by CERN EP-R\&D. 
The work is partly supported by Operational Programme Johannes Amos Comenius financed by European Structural and Investment Funds and the Czech Ministry of Education, Youth and Sports (Project LASCIMAT – CZ.02.01.01/00/23\_020/0008525). 
During the preparation of this work, the authors used Gemini (AI) in order to clarify and refine the content of the article. After using this tool, the authors reviewed and edited the content as needed and take full responsibility for the content of the publication.
\appendix

\section{Scintillation kinetics metrics}
\begin{figure*}[!ht]
    \centering
    \includegraphics[width = \linewidth]{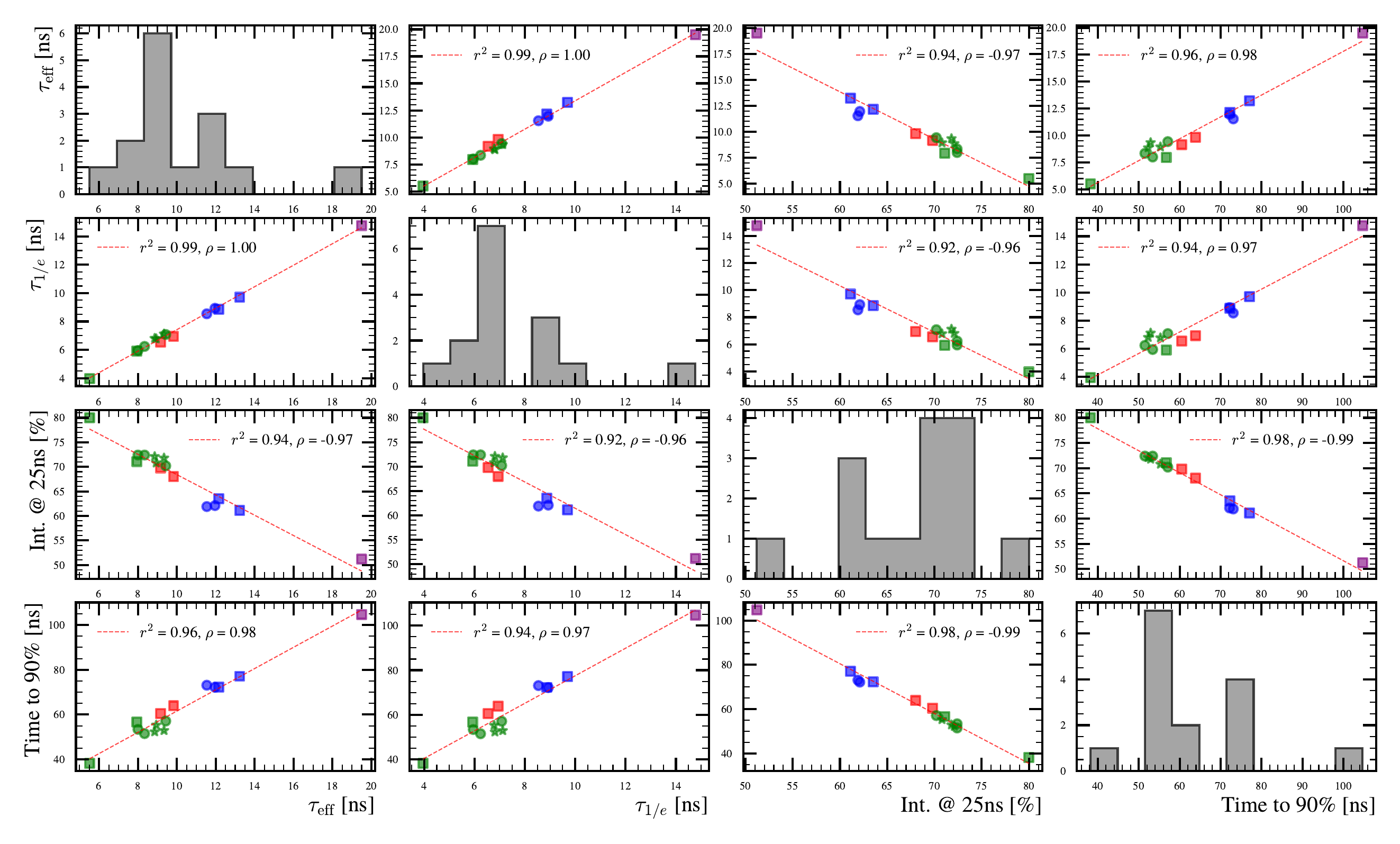}
    \caption{Correlation matrix of scintillation kinetics parameters. \textit{Blue points: ingot C, green points: ingot B, red points: ingot A and the purple point: ingot D. Circles: $2\times2\times3$ mm$^3$, stars: $2\times2\times10$ mm$^3$ and squares represent plates.}}
    \label{fig:corr_matrix}
\end{figure*}

To further explore the relationships among these new parameters, aiming at assessing how fast the scintillation pulse is and how well its light is contained within 25 ns, we provide a correlation matrix of the four parameters introduced in this article ($\tau_{\mathrm{d,eff}}$, $\tau_{1/e}$, Time for 90\%, Integral at 25 ns). This analysis aims to better understand the underlying connections between them, as illustrated in Fig.~\ref{fig:corr_matrix}. 

For each subplot, the coefficient of determination $r^2$ and the Pearson correlation coefficient $\rho$ are given in the legend.

\bibliographystyle{cas-model2-names}

\bibliography{references}

\end{document}